\documentclass[journal]{IEEEtran}
\ifCLASSINFOpdf

\else

\fi

\usepackage[cmex10]{amsmath}
\usepackage{amssymb}
\usepackage{cite}
\usepackage{graphicx}
\usepackage{array,color}
\usepackage{amsmath}
\usepackage{stfloats}
\usepackage{graphicx}
\usepackage{subfigure}
\usepackage{tabularx}
\usepackage{epsfig,epsf,color,balance,cite}
\usepackage{algorithmic}
\usepackage{algorithm}
\usepackage{bm}
\usepackage{textcomp}
\usepackage[justification=centering]{caption}
\usepackage{cuted}

\begin{document}
\title{Secure Wireless Communication in Active RIS-Assisted DFRC System}
\author{Yang Zhang, Hong Ren, \textit{Member IEEE}, Cunhua Pan, \textit{Senior Member IEEE},\\ Boshi Wang, Zhiyuan Yu, Ruisong Weng, Tuo Wu, Yongchao He}

\markboth{Journal of \LaTeX\ Class Files,~Vol.~00, No.~00, Feburary~2024}%
{Secure Wireless Communication in Active RIS-Assisted MISO DFRC Systems}

\maketitle
\vspace{-1.5cm}

\let\thefootnote\relax\footnotetext{Y. Zhang, H. Ren, C. Pan, B. Wang, Z. Yu, R. Weng and Y. He are with National Mobile Communications Research Laboratory, Southeast University, Nanjing, China. (email: {220230982, hren, cpan, 220230800, zyyu, 230239439, heyongchao}@seu.edu.cn). T. Wu is with the School of Electronic Engineering and Computer Science, Queen Mary University of London, E1 4NS London, U.K. (e-mail: tuo.wu@qmul.ac.uk).

\textit{Corresponding authors: Hong Ren and Cunhua Pan.}}

\begin{abstract}
This work considers a dual-functional radar and communication (DFRC) system with an active reconfigurable intelligent surface (RIS) and a potential eavesdropper. 
Our purpose is to maximize the secrecy rate (SR) of the system by jointly designing the beamforming matrix at the DFRC base station (BS) and the reflecting coefficients at the active RIS, subject to the signal-to-interference-plus-noise-ratio (SINR) constraint of the radar echo and the power consumption constraints at the DFRC-BS and active RIS. 
An alternating optimization (AO) algorithm based on semi-definite relaxation (SDR) and majorization-minimization (MM) is applied to solve the SR-maximization problem by alternately optimizing the beamforming matrix and the reflecting coefficients. 
Specifically, we first apply the SDR and successive convex approximation (SCA) methods to transform the two subproblems into more tractable forms, then the MM method is applied to derive a concave surrogate function and iteratively solve the subproblems.
Finally, simulation results indicate that the active RIS can better confront the impact of ``multiplicative fading" and outperforms traditional passive RIS in terms of both secure data rate and radar sensing performance.
\end{abstract}

\IEEEpeerreviewmaketitle

\section{Introduction}\label{intrdc}
\IEEEPARstart{I}{ntegrated} sensing and communication (ISAC) has attracted increasing attention in recent years as one of the key candidate technologies to provide high-quality wireless communication and radar sensing function.
However, the sensing function
requires extra bandwidth and thereby intensifying the problem of spectrum congestion.
As a promising solution, the  dual-functional radar and communication (DFRC) is regarded as the paradigm of ISAC network for its resource efficiency, which is capable of realizing communication and radar spectrum sharing (CRSS) as well as the shared use of hardware platform\cite{zheng2019radar, he2022ris, liu2018toward, liu2018dual, hassanien2015dual}.

However, the dual-functional nature of DFRC signal may lead to serious security concerns which is often ignored in relevant studies.
It is noteworthy that the DFRC transmit signal is a jointly-designed waveform carrying both communication and sensing signal. 
Hence, the embedded data can be easily leaked to potential eavesdroppers and even targets when there is no specific design to deal with the security issue.
There have been several studies in secure wireless communication in DFRC systems recently \cite{deligiannis2018secrecy, chalise2018performance, su2020secure, su2022secure}. 
In particular, the authors of \cite{su2020secure, su2022secure} considered the scenario where the targets serve as the potential eavesdroppers in the multiple-input multiple-output (MIMO) DFRC system, and proposed various algorithms for the cases of imperfect, statistical channel state information (CSI), and uncertain target direction estimation.
Nevertheless, when the channel of the legitimate communication link and that of the eavesdropping link are highly correlated, the aforementioned techniques face limitations in terms of secrecy rate (SR).

Reconfigurable intelligent surfaces (RISs) can be an effective solution to the aforementioned problems owing to its capacity of manipulating the wireless environment.
Specifically, an RIS is a meta-surface consisting of various low-cost passive reflecting elements, each of which is capable of controlling the phase shift of incident signal independently.
For communication tasks, the RIS can establish virtual links and reconfigure the reflected signals constructively for legitimate users and destructively for eavesdroppers by controlling the elements.
Consequently, it can strengthen the quality of service (QoS) of legitimate users and weaken that of eavesdroppers, and thus enhance the secure communication performance of the DFRC system\cite{chen2019intelligent,chu2019intelligent,cui2019secure,yu2021irs,xu2019resource,zhou2021secure}.
For sensing tasks, the deployment of RIS can create a virtual line-of-sight (LoS) link for the sensing targets located in the non-line-of-sight (NLoS) area of the DFRC base station (BS) and solve the problem that the millimeter wave (MMW) can be easily obstructed. Therefore,  the deployment of RIS is regarded as an appealing innovation in ISAC systems.
Thanks to the above-mentioned advantages of the RIS, transmit design for RIS-assisted DFRC systems have been well studied over the past few years\cite{fang2021sinr,jiang2021intelligent, song2022joint, sankar2022beamforming, he2022ris, WQQjoint}.
Generally, transmit beamforming vectors and the RIS coefficients are jointly optimized in these studies for desired performance gain in communication or radar sensing.
In particular, the authors of {\cite{ fang2021sinr }} considered the signal-to-interference-and-noise-ratio (SINR) maximization for an RIS-assisted secure DFRC system.
However, their system model requires extra complexity of signal processing and complicated hardware design to evaluate the radar SINR.
In contrast, the authors of \cite{jiang2021intelligent} considered the four-hop echo model which is more practical in RIS-assisted DFRC systems and proposed some methods to tackle the quartic expression of the four-hop radar SINR without considering the security problem.

Nevertheless, the performance gain of these passive RIS-assisted schemes is limited owing to the “multiplicative fading” effect of the RIS-reflecting link, i.e., the equivalent path loss of the multi-hop link is the product of those of the discrete links.
As demonstrated in previous works, the capacity gain provided by the RIS-reflecting channel can only be observed when the direct link is very weak due to the severe equivalent path loss\cite{zhang2022active}.
This may result in limited QoS at legitimate users and low received signal power at the BS, limiting the dual-functional performance of the system.
Although passive RIS can provide an array gain proportional to square of the number of its elements\cite{Wu2021Intelligent,PAN2022overview}, it requires a large number of reflecting elements to overcome the severe propagation loss, which will result in high channel estimation overhead and excessive optimization complexity\cite{hu2021two, pan2020multicell}.

To overcome the “multiplicative fading” effect in passive RIS-assisted systems, the novel concept of active RIS has been proposed recently\cite{zhang2022active}.
The key innovation of the active RIS is that each reflecting element is equipped with a reflection-type amplifier, so the amplitude and phase shift of the reflected signal can be adjusted simultaneously with extra power consumption and non-negligible thermal noise.
Based on the amplification function, the active RIS can efficiently address the severe fading issue and acquire higher received signal power through the RIS-reflecting channel compared with the traditional passive RIS\cite{zhi2022active}.
In terms of the security of wireless networks, the active RIS can achieve much better QoS of legitimate users, while destructively reconfiguring the signals to the eavesdroppers for less data leakage. Hence, active RISs  have been recently amalgamated with physical layer security. 
For example, the authors of \cite{dong2021active} and \cite{lv2022green} have proposed a novel scheme of active RIS-assisted secure communication in a multiple-input single-output (MISO) system, and presented solutions to several optimization problems including SR maximization.
However, all these works are restricted to non-sensing settings and their algorithms cannot be applied straightforward to the DFRC models.

Owing to the aforementioned research gaps, the deployment of the active RIS is also regarded as a promising innovation in secure DFRC systems.
Except for the advantages in security concerns, it is also demonstrated that the active RIS is suitable to overcome the severe propagation loss of the multi-hop radar echo channel, and obtain higher signal power at the receiver of the DFRC-BS for better sensing resolution\cite{mylonopoulos2022active,zhang2022activeCRAN, yu2023activeris}.
However, there are only a few contributions devoted to the security problem in active RIS-assisted DFRC systems. In particular, the authors of \cite{salem2022active} studied the SR maximization problem in an active RIS-assisted mutiple-user multiple-input single-output (MU-MISO) DFRC system by jointly designing the transmit beamformers, the radar receive beamformers and the coefficient matrix at the active RIS. They also took four-hop radar SINR to evaluate the sensing function of the DFRC system. However, the echo noise and interference received by the DFRC-BS and extra power consumption were totally neglected in their system model, which is an ideal assumption that ignore the practical features of the active RIS.

To the best of our knowledge, this paper is the first work to study the SR maximization problem in an MISO DFRC system while considering the non-negligible interference and noise in the four-hop radar echo signal as well as the power consumption at the active RIS.
The main contributions of this paper are summarized as follows:

\begin{itemize}
	\item[1)]
	We consider an active RIS-assisted DFRC system with a DFRC-BS, a single-antenna legitimate user and a potential eavesdropper as well as a target located in the NLoS area of the BS.
	Then, we formulate an SR maximization problem subject to the transmit power budget and radar SINR constraints and optimize the BS precoding matrix and the RIS reflecting coefficients. 
\end{itemize}

\begin{itemize}
	\item[2)]
	We propose an alternating optimization (AO) algorithm based on the semi-definite relaxation (SDR) and majorization-minimization (MM) algorithm to tackle the non-convex problem.
    In each subproblem decomposed by the AO algorithm, we first use the method of successive convex approximation (SCA) to handle the radar SINR constraint and apply SDR to reformulate the problem. Then we obtain the concave approximation function by using the first-order Taylor expansion, and apply the MM algorithm to iteratively solve the subproblems. By alternately optimizing the precoding matrix and the reflecting coefficients, the algorithm can converge to a sub-optimal point of the original problem.
\end{itemize}

\begin{itemize}
	\item[3)]
    Simulation results are presented to verify the advantages of the active RISs over traditional passive RISs in our proposed scenario.
	It is demonstrated that both the secrecy rate and the radar SINR of the DFRC system can be significantly enhanced with the deployment of an active RIS. 
	Besides, the convergence behavior of the SDR-MM-based AO algorithm are verified.
\end{itemize}

The remainder of this paper is organized as follows. In Section \uppercase\expandafter{\romannumeral+2}, we present the system model of the active RIS-assisted DFRC system and formulate the SR maximization problem. 
In Section \uppercase\expandafter{\romannumeral+3}, we develop an SDR-MM-based algorithm. 
Simulation results are provided in Section \uppercase\expandafter{\romannumeral+4}, and Section \uppercase\expandafter{\romannumeral+5} briefly concludes the paper.

\emph{Notations}: Constants, vectors and matrices are denoted by italics, boldface lowercase and boldface uppercase letters, respectively. For a complex value $a$, ${\rm{Re}}\left (a \right )$ denotes the real part of $a$. $[b]^{+}$ denotes ${\rm{max}}(b, 0)$. ${{\mathbb{C}}^{M \times N }}$ denotes the set of $M \times N$ complex vectors or matrices. ${{\mathbb{E}}}\{\cdot\} $ denotes the expectation operation. ${\left\| {\bf{x}} \right\|}_{2}$ denotes the 2-norm of vector ${\bf{x}}$. ${\left\| {\bf{A}} \right\|_{\rm{F}}}$ and ${\rm{tr}}\left( {\bf{A}} \right)$ denote the Frobenius norm and trace of ${\bf{A}}$, respectively. ${ \bf{A} ^{*}}$, ${ \bf{A}^{\rm{T}}}$ and ${  \bf{A} ^{\rm{H}}}$ denote the conjugate, transpose and Hermitian transpose of $\bf{A}$, respectively. ${\bf{A}}_{[m,n]}$ denotes the $\left ( m, n \right ) $-th entry of ${\bf{A}}$. ${\rm{diag}}(\cdot)$ and ${\rm{vec}}(\cdot)$ represent the diagonalization and vectorization operators, respectively. $\Sigma\left({\bf{x}} \right) $ denotes the operator of transforming vector ${\bf{x}} \in {\mathbb{C}}^{N^{2} \times 1}$ into an ${N \times N}$ matrix. $\nabla{f}\left( {\bf{x}} \right)$ denotes the gradient of the function $f$ with respect to (w.r.t.) the vector ${\bf{x}} $. ${\bf{B}} \otimes {\bf{C}}$ denotes the Kronecker product of ${\bf{B}}$ and ${\bf{C}}$. ${\bf{I}}_N$ denotes the $N \times N$ identity matrix, and ${\bf{r}} \sim {\cal C}{\cal N}({\bf{0}},{\bf{I}})$ denotes a random vector following the Gaussian distribution of zero mean and unit variance.

\begin{figure}[h]
	\centering
	\includegraphics[width=3in]{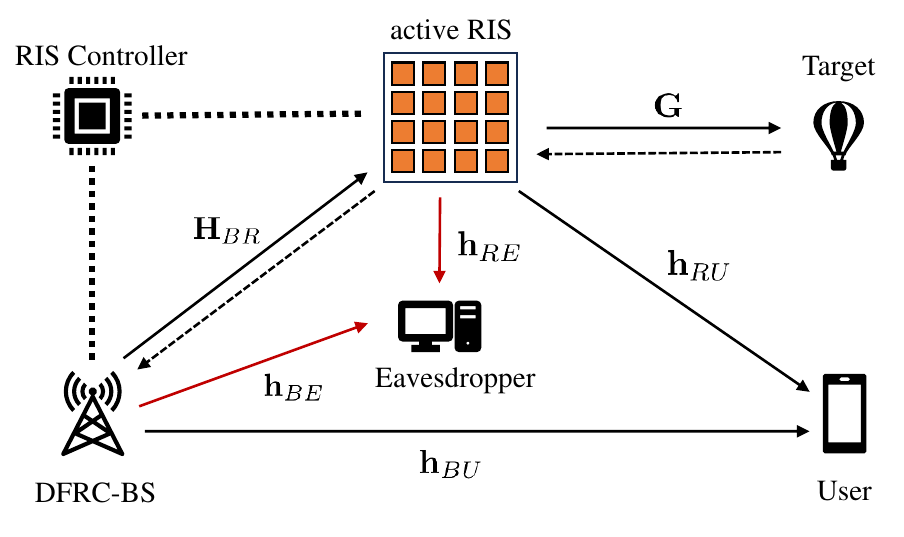}
	\caption{An active RIS-assisted secure DFRC system.}
	\vspace{-0.3cm}
	\label{fig1}
\end{figure}

\section{System Model and Problem Formulation}\label{system}
\subsection{System Model}\label{sysmod}
As shown in Fig. 1, we consider an active RIS-assisted secure DFRC system. The DFRC-BS equipped with a uniform linear array (ULA) of $M$ antennas  serves a single-antenna legitimate user while a single-antenna eavesdropper intends to wiretap the confidential information.
Meanwhile, the DFRC-BS also intends to detect a point-like target, which is located at the NLoS area of the DFRC-BS.
Besides, an active RIS with $N$ reflecting elements is deployed in the system to assist the radar sensing and secure communication.

\subsubsection{Transmit Signal Model}\label{sigmod}
We consider that the BS sends both confidential communication and dedicated sensing signals using transmit beamforming technique to facilitate the dual function of the system. Let us denote $s$ as the confidential communication signal and ${\bf{d}} \in {\mathbb{C}}^{M \times 1} $ as the dedicated sensing signal, where the signals $s$ and ${\bf{d}}$ are independent of each other, and satisfy $s \sim {\cal C}{\cal N}(0,1)$, ${\mathbb{E}}\left \{ {\bf{d}} \right \} = {\bf{0}}$ and ${\mathbb{E}}\left \{ {\bf{d}}{\bf{d}}^{\rm{H}} \right \}= {\bf{I}}_M$. 
We denote ${\bf{W}}_{\rm{r}} \triangleq \left [{\bf{w}}_{{\rm{r}},1} ,{\bf{w}}_{{\rm{r}},2},\cdots,{\bf{w}}_{{\rm{r}},M} \right ]  \in {\mathbb{C}}^{M \times M}$ and ${\bf{w}}_{\rm{c}} \in {\mathbb{C}}^{M \times 1}$ as the beamforming matrix and vector of sensing and communication signals, respectively. Therefore, the transmit signal of the DFRC-BS is given by
\begin{equation}\label{aaaaaa}
	{\bf{x}} = {\bf{W}}_{\rm{r}}{\bf{d}}+{\bf{w}}_{\rm{c}}s \triangleq {\bf{W}}{\hat{\bf{s}}},
\end{equation}
where ${\bf{W}} \triangleq \left[ {\bf{W}}_{\rm{r}}, {\bf{w}}_{\rm{c}} \right] \in {\mathbb{C}}^{M\times\left ( M+1 \right ) }  $, ${\hat{\bf{s}}} \triangleq \left[{\bf{d}}^{\rm{T}}, s \right]^{\rm{T}} \in {\mathbb{C}}^{\left( M+1\right) \times 1} $, and the covariance matrix of the transmit signal can be written as 
\begin{equation}\label{aaaaal}
	{\bf{R}} = {\mathbb{E}}\left \{ {\bf{x}}{\bf{x}}^{\rm{H}} \right \} = {\bf{W}}{\bf{W}}^{\rm{H}}  = \sum_{i=1}^{M+1}{\bf{w}}_{i}{\bf{w}}_{i}^{\rm{H}},
\end{equation}
where ${\bf{w}}_i$ denotes the $i$-th column of matrix ${\bf{W}}$.
We first consider the secure communication function and denote ${{\bf{H}}_{\rm{BR}}} \in \mathbb{C}^{N\times M}$, ${{\bf{h}}_{\rm{BU}}} \in \mathbb{C}^{M\times 1}$, ${{\bf{h}}_{\rm{BE}}} \in \mathbb{C}^{M\times 1}$, ${{\bf{h}}_{\rm{RU}}} \in \mathbb{C}^{N\times 1}$, and ${{\bf{h}}_{\rm{RE}}} \in \mathbb{C}^{N\times 1}$ as the channel coefficients from the BS to the RIS, from the BS to the user, from the BS to the eavesdropper, from the RIS to the user and from the RIS to the eavesdropper, respectively.
We also assume that the CSI of the above channels is perfectly known at the DFRC-BS by applying low-complexity channel estimation methods\cite{wang2020channel, guan2021anchor,ZHOU2022CHANNEL}. 
It is assumed that the eavesdropper can actively attack the system. 
Specifically, it can misdirect the DFRC-BS and pretend to be a legitimate user by sending the pilot signals to the DFRC-BS during the channel estimation phase{\cite{zhu2014secure}}.
Furthermore, owing to the severe multiplicative path loss, we can neglect the signals reflected by the active RIS for more than two times.
Therefore, the signal received at the legitimate user and the eavesdropper can be respectively given by
\begin{equation}\label{aaaaab}
	{{{y}}_{\rm{U}}} = \left ( {\bf{h}}_{\rm{BU}}^{\rm{H}} + {\bf{h}}_{\rm{RU}}^{\rm{H}}{\bf\Phi}  {\bf{H}}_{\rm{BR}}  \right ){\bf{x}} + {\bf{h}}_{\rm{RU}}^{\rm{H}} {\bf\Phi}{{\bf{v}}_{1}}+{{n}}_{\rm{U}},
\end{equation}
\begin{equation}\label{aaaaac}
	{{{y}}_{\rm{E}}} = \left ( {\bf{h}}_{\rm{BE}}^{\rm{H}} + {\bf{h}}_{\rm{RE}}^{\rm{H}}  {\bf\Phi}  {\bf{H}}_{\rm{BR}}  \right ){\bf{x}} + {\bf{h}}_{\rm{RE}}^{\rm{H}} {\bf\Phi}{{\bf{v}}_{1}}+{{n}}_{\rm{E}},
\end{equation}
where  $n_{\rm{U}}\sim {\cal C}{\cal N}\left( 0,\sigma^2_{\rm{U}} \right)$ and $n_{\rm{E}}\sim {\cal C}{\cal N}\left( 0,\sigma^2_{\rm{E}} \right)$ denote the additive white Gaussian noise (AWGN) at the user and eavesdropper, respectively, and matrix ${\bf{\Phi}} ={\rm{diag}}\left (\phi _{1},\phi _{2}, \dots ,  \phi_{N}\right )$ is the diagonal reflecting coefficient matrix of the active RIS with $\phi_i = \beta _{i}e^{j\vartheta _{i}}$, where $\beta _{i}\ge0$ and $\vartheta _{i}\in \left[0,2\pi\right)$ are the amplitude and the phase shift of the $i$-th reflecting element, respectively. Compared with traditional passive RISs, the elements of an active RIS are equipped with reflection-type amplifiers which consume extra power to amplify the incident signal, and thus the thermal noise at the active RIS is non-negligible. Furthermore, the amplitude $\beta _{i}$ also represents the amplification gain of the $i$-th reflecting element, and satisfies  $\beta _{i} \le \eta_i$ with $\eta_i > 1$  denoting the maximum amplification gain. Finally, the vector ${\bf{v}}_1 \sim  {\cal C}{\cal N}\left( {\bf{0}},\sigma^2_{v_1}{\bf{I}}_{N} \right)$ denotes the zero-mean AWGN at the active RIS.

Then, we consider the sensing function of the DFRC system. We assume that the potential target is located in the NLoS area of the BS owing to the blockage, and the virtual LoS channel created by the active RIS is much stronger than the NLoS one. Hence, the effect of NLoS channel between the BS and sensing target can be neglected. In addition, it is assumed that the location of the active RIS is well designed with few obstacles. Therefore, we define the response channel matrix ${\bf{G}}\in \mathbb{C}^{N\times N}$ of the sensing target to the RIS as
\begin{equation}
	{\bf{G}} = \gamma {\bf{a}}\left ( \theta  \right ) {\bf{a}}^{\rm{H}}\left ( \theta  \right ),
\end{equation}
which is based on the clutter-free model{\cite{sun2015mimo, sun2015waveform}}, where $\gamma$ denotes the complex path loss coefficient of the response channel. The vector ${\bf{a}}\left ( \theta  \right )$ represents the steering vector of the RIS and $\theta$ denotes the angle of departure (AoD) of the target towards the active RIS, which is given by
\begin{equation}\label{aaaaad}
		{\bf{a}}\left ( {\theta} \right )  = \left [ 1, e^{j2\pi \frac{d_{\rm{RIS}}}{\lambda}  {\rm{sin}} {\theta}  } ,\dots, e^{j2\pi \frac{d_{\rm{RIS}}}{\lambda} \left ( N-1 \right ) {\rm{sin}} {\theta} } \right ]^{\rm{H}},
\end{equation}
where $d_{\rm{RIS}}$ denotes the interval between adjacent RIS elements and $\lambda$ denotes the carrier wavelength.
Since the information signal $s$ and the dedicated sensing signal ${\bf{d}}$ can be jointly utilized to illuminate the sensing target, the radar echo received at the DFRC-BS can be given by ({\ref{aaaaae}}) at the bottom of the next page,
\begin{figure*}[hb]
\vspace{-0.3cm}
{\noindent}	 \rule[5pt]{18cm}{0.1em}\\
\vspace{-0.3cm}
\begin{equation}\label{aaaaae}
		{\bf{y}}_{\rm{R}} = {\bf{H}}^{\rm{H}}_{\rm{BR}}{\bf\Phi}^{\rm{H}} {\bf{G}}{\bf\Phi}  {\bf{H}}_{\rm{BR}}{\bf{x}} + {\bf{H}}^{\rm{H}}_{\rm{BR}}{\bf\Phi}^{\rm{H}} {\bf{G}}{\bf\Phi}{{\bf{v}}_{1}}+{\bf{H}}^{\rm{H}}_{\rm{BR}}{\bf\Phi}{{\bf{v}}_{1}}+{\bf{H}}^{\rm{H}}_{\rm{BR}}{\bf\Phi}^{\rm{H}} {{\bf{v}}_{2}}+\rho{\bf{H}}^{\rm{H}}_{\rm{BR}}{\bf\Phi}  {\bf{H}}_{\rm{BR}}{\bf{x}} +{\bf{n}}_{\rm{R}}. 
\end{equation}
\vspace{-0.7cm}
\end{figure*}
where the vectors ${\bf{v}}_2 \sim {\cal C}{\cal N} \left( {\bf{0}},\sigma^2_{v_2}{\bf{I}}_{N} \right)$ and ${\bf{n}}_{\rm{R}} \sim {\cal C}{\cal N} \left( {\bf{0}},\sigma^2_{\rm{R}}{\bf{I}}_{M} \right)$ denote the dynamic thermal noise at the active RIS in the uplink reflected signal and the noise at the DFRC-BS receiver, respectively. 
Note that the echo directly reflected by the RIS in the first reflection in (\ref{aaaaae}), namely ${\bf{H}}^{\rm{H}}_{\rm{BR}}{\bf\Phi}  {\bf{H}}_{\rm{BR}}{\bf{x}}$, is considered as interference with  no information of the sensing target.
By adopting effective self-interference (SI) cancellation techniques\cite{yu2023activeris}, the influence of interference signal at the DFRC-BS can be reduced to some extent,  where ${\rho}$ denotes the SI coefficient after mitigation.

Finally, the active RIS first reflects the transmit signal to the legitimate user, eavesdropper and the target, then reflects the echo signal from the target to the DFRC-BS.
The first and second reflected signal in this process can be respectively given by 
\begin{align}
	{\bf{y}}_{\rm{r_1}} &= {\bf\Phi}  {\bf{H}}_{\rm{BR}}{\bf{x}} + {\bf\Phi}{{\bf{v}}_{1}},\label{aaaaaf}\\
	{\bf{y}}_{\rm{r_2}} &= {\bf\Phi}^{\rm{H}} {\bf{G}}{\bf\Phi}  {\bf{H}}_{\rm{BR}}{\bf{x}} + {\bf\Phi}^{\rm{H}} {\bf{G}}{\bf\Phi}{{\bf{v}}_{1}}+{\bf\Phi}^{\rm{H}} {{\bf{v}}_{2}}.\label{aaaaag}
\end{align}

\subsubsection{Metrics of Dual Functional Performance}\label{metrics}
Firstly, based on the aforementioned model of transmit signal, the signal-to-noise-ratio (SNR) of the legitimate user and the eavesdropper are given in (\ref{aaaaaw}) and (\ref{aaaaaq}), respectively.
\begin{equation}\label{aaaaaw}
	\xi_{\rm{U}} =  \frac{{\left |   \left ( {\bf{h}}_{\rm{BU}}^{\rm{H}} + {\bf{h}}_{\rm{RU}}^{\rm{H}}  {\bf\Phi}  {\bf{H}}_{\rm{BR}}  \right ){\bf{w}}_{\rm{c}}\right |}^{2}}{\sigma^2_{v_1}{\left \| {\bf{h}}_{\rm{RU}}^{\rm{H}} {\bf\Phi} \right \|}^{2}_{2} +\sigma^2_{\rm{U}}}, 
\end{equation}
\begin{equation}\label{aaaaaq}
	\xi_{\rm{E}} =  \frac{{\left |   \left ( {\bf{h}}_{\rm{BE}}^{\rm{H}} + {\bf{h}}_{\rm{RE}}^{\rm{H}}  {\bf\Phi}  {\bf{H}}_{\rm{BR}}  \right ){\bf{w}}_{\rm{c}}\right |}^{2}}{\sigma^2_{v_1}{\left \| {\bf{h}}_{\rm{RE}}^{\rm{H}} {\bf\Phi} \right \|}^{2}_{2} +\sigma^2_{\rm{E}}}.
\end{equation}
Note that the dedicated sensing signal ${{\bf{W}}_{\rm{r}}{\bf{d}}}$ can be generated offline and known by the user as well as the eavesdropper prior to the transmission, so both the legitimate user and the eavesdropper can cancel the interference generated by the sensing signal ${{\bf{W}}_{\rm{r}}{\bf{d}}}$ with its prior knowledge\cite{song2022joint}.
Therefore, the achievable transmission rate (nat/s/Hz) of the user and eavesdropper can be respectively written as 
\begin{equation}\label{alskwo}
	R_{\rm{U}} = \ln {\left ( {1+{\xi_{\rm{U}}}}  \right ) }, R_{\rm{E}} = \ln {\left ( {1+{\xi_{\rm{E}}}}  \right ) }. 
\end{equation}

Then, the secrecy rate from the user to the DFRC-BS is given by
\begin{equation}\label{aaaaas}
	S =  \left [ R_{\rm{U}}-R_{\rm{E}} \right ] ^{+}.
\end{equation}

For radar sensing function, it is worth noting that the DFRC-BS receiver has complete knowledge of the transmit signal ${\bf{x}}$, so the communication waveform as well as the sensing signal in the radar echo can be utilized. Let us define the following matrices 
\begin{equation}\nonumber
	{\bf{A}} \triangleq {\bf{H}}^{\rm{H}}_{\rm{BR}}{\bf\Phi}^{\rm{H}} {\bf{G}}{\bf\Phi}{\bf{H}}_{\rm{BR}}, {\bf{B}} \triangleq {\rho}{\bf{H}}^{\rm{H}}_{\rm{BR}}{\bf\Phi}{\bf{H}}_{\rm{BR}},
\end{equation}
\begin{equation}\nonumber
    {\bf{C}} \triangleq {\bf{H}}^{\rm{H}}_{\rm{BR}}{\bf\Phi}^{\rm{H}} {\bf{G}}{\bf\Phi}+ {\bf{H}}^{\rm{H}}_{\rm{BR}}{\bf\Phi}, {\bf{D}}\triangleq{\bf{H}}_{\rm{BR}}^{\rm{H}}{\bf\Phi}^{\rm{H}},
\end{equation}
then the SINR of radar echo received at the DFRC-BS receiver can be written as\cite{WQQjoint},\cite{LIBORADAR}
\begin{equation}\label{abaerf}
	\xi_{\rm{R}} = {\rm{tr}}\left ( {{\bf{A}}   {\bf{R}}  {\bf{A}}^{\rm{H}}}  {\bf{J}} ^{-1}  \right ),
\end{equation}
where ${\bf{J}}$ represents the interference-plus-noise covariance matrix, which is given by\cite{zheng2017joint}
\begin{equation}\label{awpslf}
		{\bf{J}}={{\bf{B}}{\bf{R}}{\bf{B}}}^{\rm{H}}+{\bf{N}},
\end{equation}
where
\begin{equation}\label{nshism}
	{\bf{N}}=\sigma_{v_1}^{2}{\bf{C}}{\bf{C}}^{\rm{H}} +\sigma_{v_2}^{2} {\bf{D}}{\bf{D}}^{\rm{H}}  + \sigma_{\rm{R}}^{2}{\bf{I}}_{M}.
\end{equation}

In addition, the transmit power of the DFRC-BS can be expressed as
\begin{equation}\label{acegyj}
	{P_{\rm{t}}} =  \mathbb{E}\left \{ \left \| {\bf{x}} \right \|^{2}_{2}  \right \}    = {\rm{tr}} \left( {\bf{R}} \right),
\end{equation}
and the power consumption of the active RIS in the first and second reflection can be respectively given by ({\ref{axetjm}}) and ({\ref{acholf}}) at the bottom of this page.
\begin{figure*}[hb]
\begin{equation}\label{axetjm}
		{P_{\rm{A_1}}}  = \mathbb{E} \left\{{\left \| {\bf{y}}_{\rm{r_1}} \right \| }^{2}_{2} \right\}
		 =    {\rm{tr}}\left ( {\bf{H}}_{\rm{BR}}^{\rm{H}} {\bf\Phi}^{\rm{H}}{\bf\Phi}{\bf{H}}_{\rm{BR}}{\bf{R}}\right )  +\sigma_{v_1}^{2}{\left \| {\bf\Phi}   \right \| }^{2}_{\rm{F}}.
\end{equation}
\begin{equation}\label{acholf}
		{P_{\rm{A_2}}} = \mathbb{E}\left\{{\left \| {\bf{y}}_{\rm{r_2}} \right \| }^{2}_{2} \right\}
		=    {\rm{tr}}\left ( {\bf{H}}_{\rm{BR}}^{\rm{H}}{\bf\Phi}^{\rm{H}}{\bf{G}}^{\rm{H}}{\bf\Phi}{\bf\Phi}^{\rm{H}} {\bf{G}}{\bf\Phi}  {\bf{H}}_{\rm{BR}}{\bf{R}}\right ) 
		+\sigma_{v_1}^{2}{\left \|{\bf\Phi}^{\rm{H}} {\bf{G}} {\bf\Phi}   \right \| }^{2}_{\rm{F}}+\sigma_{v_2}^{2}{\left \| {\bf\Phi}   \right \| }^{2}_{\rm{F}}.
\end{equation}
\end{figure*}

\subsection{Problem Formulation}\label{probfor}
In this paper, we aim to maximize the SR of the DFRC system by jointly optimizing the transmit beamforming matrix $\bf{W}$ at the DFRC-BS and the reflecting coefficient matrix $\bf{\Phi}$ at the active RIS, subject to the transmit power constraints at the DFRC-BS and the active RIS, the minimum radar SINR constraint and the maximum amplification gain constraint. 
Therefore, the SR maximization problem
can be formulated as
\begin{subequations}\label{ppppp1}
	\begin{align}
		\mathop {\max}\limits_{{{\bf{W}},{\bf{\Phi}} } } \quad
		&S\\
		\qquad\ \textrm{s.t.}\quad
		&  \xi_{\rm{R}} \ge {\gamma}_{\rm{r}},     \\
		&  {P_{\rm{t}}} \le P_0,  \\
		&{P_{\rm{A_1}}}+{P_{\rm{A_2}}} \le P_{\rm{RIS}},\\
		&\left |{\bf{\Phi}}_{\left [ i,i  \right ]}   \right | \le \eta_i, i = 1, 2, \dots, N,
	\end{align}
\end{subequations}
where $P_0$ is the transmit power budget of the DFRC-BS, $P_{\rm{RIS}}$ is the total power consumption budget of the active RIS, and $\gamma_{\rm{r}}$ denotes the minimum SINR threshold at the DFRC-BS receiver with given azimuth angle from the target towards active RIS.

Problem (\ref{ppppp1}) is obviously non-convex due to the complex form and highly coupled variables in the objective function and constraints. It is challenging to directly obtain an optimal solution of the original problem, so an AO algorithm is applied to solve the problem.

\section{The SDR-MM-based AO Algorithm}\label{propalgo}
In this section, we decouple the original problem into two subproblems, and alternately optimize the transmit beamforming matrix ${\bf{W}}$ and the reflecting coefficient matrix ${\bf{\Phi}}$. We use the successive convex approximation (SCA) method to tackle the radar SINR constraint, and apply an SDR-MM-based algorithm to iteratively optimize ${\bf{W}}$ and ${\bf{\Phi}}$ in the two subproblems. 

\subsection{Optimizing  ${\bf{W}}$ with Fixed ${\bf{\Phi}}$}\label{dcrsphi}
In this subsection, the transmit beamforming matrix $\bf{W}$ at the DFRC-BS is optimized with given reflecting coefficient matrix ${\bf{\Phi}}$.
Then the original problem can be reformulated as
\begin{subequations}\label{ppppp2}
	\begin{align}
		\mathop {\max }\limits_{{{\bf{W}} } } \quad &\ln \left (\frac{1+\left | \bar{\bf{h}}_{\rm{U}}^{\rm{H}}{\bf{w}}_{{M+1}} \right |^{2}} {{1+\left | \bar{\bf{h}}_{\rm{E}}^{\rm{H}}{\bf{w}}_{M+1} \right |^{2}} } \right) \\
		\qquad\ \textrm{s.t.}\quad
		&\xi_{\rm{R}} \ge \gamma_{\rm{r}},\\
		&{\rm{tr}}\left ( {\bf{R}} \right ) \le {P}_{0},\\
		& {\rm{tr}}\left ( {\bf{TR}} \right ) \le \bar{P}_{\rm{RIS}},
	\end{align}
\end{subequations}
where  ${\bar{\bf{h}}}_{\rm{U}}^{\rm{H}}, {\bar{\bf{h}}}_{\rm{E}}^{\rm{H}} ,{\bf{T}} $ and $ \bar{P}_{\rm{RIS}}$ are respectively given by
\begin{align}
     {\bar{\bf{h}}}_{\rm{U}}^{\rm{H}}&\triangleq\left ( {\bf{h}}_{\rm{BU}}^{\rm{H}} + {\bf{h}}_{\rm{RU}}^{\rm{H}}  {\bf\Phi}  {\bf{H}}_{\rm{BR}}  \right ) /\sqrt{\sigma^2_{v_1}{\left \| {\bf{h}}_{\rm{RU}}^{\rm{H}} {\bf\Phi} \right \|}^{2}_{2} +\sigma^2_{\rm{U}}},\nonumber\\
	{\bar{\bf{h}}}_{\rm{E}}^{\rm{H}} &\triangleq \left ( {\bf{h}}_{\rm{BE}}^{\rm{H}} + {\bf{h}}_{\rm{RE}}^{\rm{H}}  {\bf\Phi}  {\bf{H}}_{\rm{BR}}  \right ) /\sqrt{\sigma^2_{v_1}{\left \| {\bf{h}}_{\rm{RE}}^{\rm{H}} {\bf\Phi} \right \|}^{2}_{2} +\sigma^2_{\rm{E}}},\nonumber\\
	{\bf{T}} &\triangleq {\bf{H}}_{\rm{BR}}^{\rm{H}} {\bf\Phi}^{\rm{H}}{\bf\Phi}{\bf{H}}_{\rm{BR}}+{\bf{H}}_{\rm{BR}}^{\rm{H}}{\bf\Phi}^{\rm{H}}{\bf{G}}^{\rm{H}}{\bf\Phi}{\bf\Phi}^{\rm{H}} {\bf{G}}{\bf\Phi}  {\bf{H}}_{\rm{BR}},\nonumber\\
	\bar{P}_{\rm{RIS}} &\triangleq {P}_{\rm{RIS}} - \left( \sigma_{v_1}^{2}+\sigma_{v_2}^{2}\right) {\left \| {\bf\Phi}   \right \| }^{2}_{\rm{F}} - \sigma_{v_1}^{2}{\left \|{\bf\Phi}^{\rm{H}} {\bf{G}} {\bf\Phi}   \right \| }^{2}_{\rm{F}}.\nonumber
\end{align}

It is obvious that Problem (\ref{ppppp2}) is still non-convex owing to the forms of objective function (\ref{ppppp2}a) and radar SINR constraint (\ref{ppppp2}b). To tackle this problem, we first apply the SCA method to reformulate constraint (\ref{ppppp2}b) into a more tractable form.
By using the following Lemma 1, we can find a convex approximation of $\xi_{\rm{R}}$.

\textbf{\emph{Lemma 1}}: Given ${\bf{X}}_{(k)}, {\bf{J}}_{(k)}$ as the value of ${\bf{X}}, {\bf{J}}$ in the $k$-th iteration, we have
\begin{equation}\label{kolsdd}
	\begin{split}
		&{\rm{tr}}\left ( {{\bf{X}}^{\rm{H}}   {\bf{J}}^{-1}  {\bf{X}}}  \right )\\ \ge &2{\rm{Re}}\left ({\rm{tr}} \left (  {\bf{X}}_{(k)}^{\rm{H}}{\bf{J}}_{(k)}^{-1} {\bf{X}}\right ) \right ) - {\rm{tr}}\left ({{\bf{J}}}_{(k)}^{-1} {\bf{X}}_{(k)} {\bf{X}}_{(k)}^{\rm{H}}  {\bf{J}}_{(k)}^{-1} {\bf{J}} \right ).
	\end{split}
\end{equation}

\textbf{\emph{Proof}}: Please refer to Appendix A.$\hfill\blacksquare$

The lemma provides a lower bound of functions with the form of ${{\rm{tr}}\left ( {{\bf{X}}^{\rm{H}}   {\bf{J}}^{-1}  {\bf{X}}}  \right )}$.
Hence, by letting ${\bf{X}} = {\bf{AW}}$ in equality (\ref{kolsdd}),  $\xi_{\rm{R}}$ can be approximated as shown in ({\ref{lowbnd}}) at the bottom of this page,  
\begin{figure*}[hb]
	\vspace{-0.3cm}
	{\noindent}	 \rule[5pt]{18cm}{0.1em}\\
	\vspace{-0.3cm}
\begin{equation}\label{lowbnd}
	\begin{split}
		{\rm{tr}}\left ( {\bf{A}}   {\bf{R}}  {\bf{A}}^{\rm{H}} {\bf{J}} ^{-1}  \right ) 
		\ge & \sum_{i=1}^{M+1}2{\rm{Re}}\left(   {{\bf{w}}}_{i(k)}^{\rm{H}}   {\bf{A}}^{\rm{H}}{\bf{J}}_{(k)}^{-1}  {\bf{A}}  {\bf{w}}_{i}\right )- \sum_{i=1}^{M+1}{\rm{tr}}\left ({\bf{J}}_{(k)}^{-1} {\bf{A}} { {\bf{w}}}_{i(k)}  { {\bf{w}}}_{i(k)}^{\rm{H}} {\bf{A}}^{\rm{H}}   {\bf{J}}_{(k)}^{-1} {\bf{J}} \right )\\
		=& \sum_{i=1}^{M+1}2{\rm{Re}}\left (  { {\bf{w}}}_{i(k)}^{\rm{H}}  {\bf{A}}^{\rm{H}} {\bf{J}}_{(k)}^{-1} {\bf{A}}  {\bf{w}}_{i}\right ) - \sum_{i=1}^{M+1}{\rm{tr}}\left ({\bf{J}}_{(k)}^{-1} {\bf{A}} {\bf{R}}_{(k)} {\bf{A}}^{\rm{H}} {\bf{J}}_{(k)}^{-1}{\bf{B}}{\bf{w}}_{i}{\bf{w}}_{i}^{\rm{H}}{\bf{B}}^{\rm{H}}\right )-{\alpha}_1.
	\end{split}
\end{equation}
\vspace{-0.7cm}
\end{figure*}
where ${\bf{w}}_{i(k)}$ denotes the value of ${\bf{w}}_{i}$ in the $k$-th iteration, ${\bf{R}}_{(k)} = {\bf{w}}_{i(k)}{\bf{w}}_{i(k)}^{\rm{H}}$ and constant ${\alpha}_1 \triangleq {\rm{tr}}\left( {\bf{J}}_{(k)}^{-1} {\bf{A}} {\bf{R}}_{(k)}   {\bf{A}}^{\rm{H}}{\bf{J}}_{(k)}^{-1}{\bf{N}}\right)$. Substituting (\ref{lowbnd}) into (\ref{ppppp2}b), (\ref{ppppp2}b) can be approximated as a more tractable form written as (\ref{axwgnj}) at the bottom of this page.
\begin{figure*}[hb]
\begin{equation}\label{axwgnj}
	\sum_{i=1}^{M+1} \left \{ {\bf{w}}_{i}^{\rm{H}}{\bf{B}}^{\rm{H}} {\bf{J}}_{(k)}^{-1} {\bf{A}} {\bf{R}}_{(k)}  {\bf{A}}^{\rm{H}}   {\bf{J}}_{(k)}^{-1}{\bf{B}}{\bf{w}}_{i} -2{\rm{Re}}\left ( {{\bf{w}}}_{i(k)}^{\rm{H}}{\bf{A}}^{\rm{H}} {\bf{J}}_{(k)}^{-1} {\bf{A}}  {\bf{w}}_{i} \right ) \right \} + \alpha_1+\gamma_r \le 0.
\end{equation}
\end{figure*}

We next apply the SDR method, let
\begin{equation}\label{sdrwwi}
	{\bf{W}}_{i} = \left[{\bf{w}}_{i}^{\rm{H}}, 1 \right]^{\rm{H}}\left[{\bf{w}}_{i}^{\rm{H}}, 1 \right]  , i = 1, 2, \cdots, M+1.
\end{equation}  
By substituting (\ref{lowbnd}) and (\ref{sdrwwi}) into Problem (\ref{ppppp2}) and relaxing the rank-1 constraint, Problem (\ref{ppppp2}) can be reformulated as
\begin{subequations}\label{ppppp4}
	\begin{align}
		\mathop {\max }\limits_{\left \{ {\bf{W}}_{i} \right \}_{i=1}^{M+1} } \quad & C\left ( {\bf{W}}_{M+1} \right )  \\
		\qquad\ \textrm{s.t.}\quad
		&\sum_{i=1}^{M+1}  {\rm{tr}}\left ( {\bf{H}}_{i}{\bf{W}}_{i} \right ) + e_1 \le 0,\\
		& \sum_{i=1}^{M+1}{\rm{tr}}\left ( {\bf{W}}_{i} \right )  \le  {P}_{0}  +M+1 ,  \\
		&\sum_{i=1}^{M+1}{\rm{tr}}\left ( {\bar{\bf{T}}}{\bf{W}}_{i} \right )  \le \bar{P}_{\rm{RIS}},\\
		& {\bf{W}}_{i} \succeq {\bf{0}},\\
		& {\bf{W}}_{i\left[ M+1,M+1\right] } = 1,
	\end{align}
\end{subequations}
where $e_1 = \alpha_1+\gamma_{\rm{r}}$,  $\bar{\bf{T}} = \begin{bmatrix}
	{\bf{T}}  & {\bf{0}}_{M \times 1}\\
	{\bf{0}}_{1 \times M} & 0
\end{bmatrix}$, ${\bf{H}}_{\rm{U}} = \begin{bmatrix}
	{\bar{\bf{h}}}_{\rm{U}}{\bar{\bf{h}}}_{\rm{U}}^{\rm{H}}  & {\bf{0}}_{M \times 1}\\
	{\bf{0}}_{1 \times M} & 1
\end{bmatrix}$, ${\bf{H}}_{\rm{E}} = \begin{bmatrix}
	{\bar{\bf{h}}}_{\rm{E}}{\bar{\bf{h}}}_{\rm{E}}^{\rm{H}}  & {\bf{0}}_{M \times 1}\\
	{\bf{0}}_{1 \times M} & 1
\end{bmatrix}$, and
\begin{equation}\label{avbryu}
	{\bf{H}}_{i} = \begin{bmatrix}
		{\bf{B}}^{\rm{H}}{\bf{J}}_{(k)}^{-1}{\bf{A}}{\bf{R}}_{(k)}{\bf{A}}^{\rm{H}}{\bf{J}}_{(k)}^{-1}{\bf{B}}  & -{\bf{A}}^{\rm{H}}{\bf{J}}_{(k)}^{-1}{\bf{A}}{\bf{w}}_{i(k)}\\
		-{\bf{w}}_{i(k)}^{\rm{H}}{\bf{A}}^{\rm{H}}{\bf{J}}_{(k)}^{-1}{\bf{A}} & 0 
	\end{bmatrix}.
\end{equation}
Furthermore, denoting ${\bf{W}}_{\rm{c}} = {\bf{W}}_{M+1}$ for simplicity, we have
\begin{equation}
	C\left ( {\bf{W}}_{\rm{c}} \right ) \triangleq \ln \left ( {\rm{tr}} \left ( {\bf{H}}_{\rm{U}}  {\bf{W}}_{\rm{c}}\right ) \right)-\ln \left ({\rm{tr}} \left ( {\bf{H}}_{\rm{E}}  {\bf{W}}_{\rm{c}}\right )\right) .
\end{equation}

Since Problem (\ref{ppppp4}) is still non-convex due to the objective function, the classical interatice approach, i.e., MM algorithm
is employed to address this problem.
By applying the following Lemma 2, we can obtain a concave surrogate function of (\ref{ppppp4}a) and iteratively solve the problem with the surrogate function.

\textbf{\emph{Lemma 2}}: For any concave function $f\left ( {\bf{X}} \right )$, we have
\begin{equation}
	f\left ( {\bf{X}} \right ) \le f\left ( {\bf{X}}_{(k)} \right ) +{\rm{tr}}\left (   \nabla f\left ( {\bf{X}}_{(k)}  \right )  \left ( {\bf{X}}- {\bf{X}}_{(k)}\right ) \right ),
\end{equation}
where ${\bf{X}}_{(k)}$ is the value of ${\bf{X}}$ in the $k$-th iteration.

\textbf{\emph{Proof}}: Please refer to \cite{sun2016majorization}.$\hfill\blacksquare$

The lemma provides a linear lower bound of any concave function.
Therefore, with feasible point ${\bf{W}}_{{\rm{c}}(k)}$, the objective function in the ${\left( k+1 \right) }$-th iteration can be lower bounded by
\begin{equation}\label{ascvvb}
	\begin{split}
		C\left ( {\bf{W}}_{\rm{c}} \right ) 
		&\ge  \ln \left ({\rm{tr}} \left ( {\bf{H}}_{\rm{U}}  {\bf{W}}_{\rm{c}}\right ) \right)-\ln \left ({\rm{tr}} \left ( {\bf{H}}_{\rm{E}}  {\bf{W}}_{{\rm{c}}(k)} \right )\right)\\
		&\quad-{\rm{tr}}\left (  \frac{{\bf{H}}_{\rm{E}}}{{\rm{tr}}\left ( {\bf{H}}_{\rm{E}}{{\bf{W}}_{{\rm{c}}(k)} }\right )}  \left ( {\bf{W}}_{\rm{c}}-{{\bf{W}}_{{\rm{c}}(k)} } \right )   \right )\\
		&\triangleq {\tilde{C}}\left ( {\bf{W}}_{\rm{c}} | {\bf{W}}_{{\rm{c}}(k) } \right ).
	\end{split}
\end{equation}

It is also simple to verify that ${\tilde{C}}\left ( {\bf{W}}_{\rm{c}} | {\bf{W}}_{{\rm{c}}(k)} \right )$ satisfies the following four conditions\cite{sun2016majorization}:

1) ${\tilde{C}}\left ( {\bf{W}}_{\rm{c}} | {\bf{W}}_{{\rm{c}}(k)} \right )|_{{\bf{W}}_{\rm{c}}={\bf{W}}_{{\rm{c}}}(k)}  = {C}\left ( {\bf{W}}_{\rm{c}}\right )|_{{\bf{W}}_{\rm{c}}={\bf{W}}_{{\rm{c}}(k)}}$;

2) $\nabla {\tilde{C}}\left ( {\bf{W}}_{\rm{c}} | {\bf{W}}_{{\rm{c}}(k)} \right )|_{{\bf{W}}_{\rm{c}}={\bf{W}}_{{\rm{c}}(k)}}  = \nabla {C}\left ( {\bf{W}}_{\rm{c}}  \right )|_{{\bf{W}}_{\rm{c}}={\bf{W}}_{{\rm{c}}(k)}}$;

3) ${\tilde{C}}\left ( {{\bf{W}}}_{\rm{c}} | {\bf{W}}_{{\rm{c}}(k)}  \right )  \le {C}\left ( {{\bf{W}}}_{\rm{c}}  \right )$;

4) ${\tilde{C}}\left ( {{\bf{W}}}_{\rm{c}} | {{\bf{W}}}_{{\rm{c}}(k)}  \right ) $ is continuous in both ${\bf{W}}_{\rm{c}}$ and ${{\bf{W}}}_{{\rm{c}}(k)}$. 

Hence, ${\tilde{C}}\left ( {\bf{W}}_{\rm{c}} | {{\bf{W}}}_{{\rm{c}}(k)}\right )$ can be considered as a surrogate function of the objective function in Problem (\ref{ppppp4}). By dropping the constant terms, Problem (\ref{ppppp4}) can be approximated as
\begin{subequations}\label{ppppp5}
	\begin{align}
			\mathop {\max }\limits_{\left \{ {\bf{W}}_{i} \right \}_{i=1}^{M+1} }
		\quad &  \ln \left ( {\rm{tr}} \left ( {\bf{H}}_{\rm{U}}  {\bf{W}}_{\rm{c}}\right ) \right)-{\rm{tr}}\left (  \frac{{\bf{H}}_{\rm{E}}{\bf{W}}_{\rm{c}}}{{\rm{tr}}\left ( {\bf{H}}_{\rm{E}}{{\bf{W}}}_{{\rm{c}}(k)}\right )}      \right )\\
		\qquad\ \textrm{s.t.}\quad &{\textrm{(\ref{ppppp4}b)}}\sim{\textrm{(\ref{ppppp4}e)}}, \nonumber
	\end{align}
\end{subequations}
which is a concave maximization problem and can be directly solved via CVX solver\cite{CVX}.  Since the rank-1 constraint is omitted in Problem (\ref{ppppp5}), the original solution ${\bf{w}}_{i}$ may not be recovered through the optimal solution. Therefore, the method of eigenvalue decomposition and Gaussian randomization are applied to construct the rank-1 solution.

According to the key property of MM algorithm, the convergence of optimizing ${\bf{W}}_i$ is non-decreasing and we can finally obtain a sub-optimal point of Problem (\ref{ppppp4}). By using the eigenvalue decomposition and Gaussian randomization methods, an optimal rank-1 solution ${\bf{W}}_{i}^{\rm{opt}}$ can be obtained, then the optimal transmit beamforming vectors ${\bf{w}}_{i}^{\rm{opt}}$ can be written as $\left( {\bm{\nu}}_{\rm{max}}\left ( {\bf{W}}_{i}^{\rm{opt}} \right )_{\left[ 1:M\right]}\right)  \sqrt{\lambda_{\rm{max}}  \left ( {\bf{W}}_{i}^{\rm{opt}} \right )} $ and ${\bf{W}}^{\rm{opt}} = \left[{\bf{w}}_{1}^{\rm{opt}}, {\bf{w}}_{2}^{\rm{opt}}, \cdots, {\bf{w}}_{M+1}^{\rm{opt}} \right] $, where ${\lambda_{\rm{max}}  \left ( {\bf{W}}_{i}^{\rm{opt}} \right )}$ and ${\bm{\nu}}_{\rm{max}}\left ( {\bf{W}}_{i}^{\rm{opt}} \right )$ denote the largest eigenvalue and its corresponding eigenvector of ${{\bf{W}}_{i}^{\rm{opt}}}$, respectively.
\subsection{Optimizing ${\bf{\Phi}}$ with Fixed ${\bf{W}}$}\label{dphiscr}
In this subsection, the reflecting  coefficient matrix $\bf{\Phi}$ at the active RIS is optimized with given beamforming matrix  ${\bf{W}}$. Then, Problem (\ref{ppppp1}) can be recast as
\begin{subequations}\label{ppppp3}
	\begin{align}
		\mathop {\max }\limits_{{{\bf{\Phi}} } } \quad & R_{\rm{U}}-R_{\rm{E}} \\
		\qquad\ \textrm{s.t.}\quad
		& \xi_{\rm{R}} \ge \gamma_{\rm{r}},\\
		&{P_{\rm{A_1}}}+{P_{\rm{A_2}}} \le P_{\rm{RIS}},\\
		&\left |{\bf{\Phi}}_{\left [ i,i  \right ]}   \right | \le \eta_i, i = 1, 2, \dots, N.
	\end{align}
\end{subequations}

Problem (\ref{ppppp3}) is still non-convex and difficult to handle due to the objective function and the quartic form of ${\bf{\Phi}}$ in constraint (\ref{ppppp3}b) and (\ref{ppppp3}c). To this end, we transform the problem into a more tractable form and apply the SDR-MM-based algorithm. 

We first denote ${\bf{v}} = \left [ \phi_1,\phi_2, \cdots ,\phi_N \right ] ^{\rm{H}} \in {\mathbb{C}}^{N \times 1}$ for further manipulations. Using vector ${\bf{v}}$, $R_i$ in the objective function can be expressed as (\ref{CICICI}) at the bottom of this page.
\begin{figure*}[hb]
	    \vspace{-0.3cm}
		{\noindent}	 \rule[5pt]{18cm}{0.1em}\\
		\vspace{-0.3cm}
	\begin{equation}\label{CICICI}
		\begin{split}
			R_{i} =& \ln  \left({\left |   \left ( {\bf{h}}_{{\rm{B}}i}^{\rm{H}} + {\bf{h}}_{{\rm{R}}i}^{\rm{H}}  {\bf\Phi}  {\bf{H}}_{\rm{BR}}  \right ){\bf{w}}_{\rm{c}}\right |}^{2} + {\sigma^2_{v_1}{\left \| {\bf{h}}_{{\rm{R}}i}^{\rm{H}} {\bf\Phi} \right \|}^{2}_{2} + \sigma^2_{i}} \right ) - \ln \left ( {\sigma^2_{v_1}{\left \| {\bf{h}}_{{\rm{R}}i}^{\rm{H}} {\bf\Phi} \right \|}^{2}_{2} +\sigma^2_{i}} \right ) \\
			=& \ln \left ({\left |   \left ( {\bf{h}}_{{\rm{B}}i}^{\rm{H}} + {\bf{v}}^{\rm{H}}{\rm{diag}}\left({\bf{h}}_{{\rm{R}}i}^{\rm{H}} \right)  {\bf{H}}_{\rm{BR}}  \right ){\bf{w}}_{\rm{c}}\right |}^{2}+ {\sigma^2_{v_1}{\left \| {\bf{v}}^{\rm{H}}{\rm{diag}}\left({\bf{h}}_{{\rm{R}}i}^{\rm{H}} \right) \right \|}^{2}_{2} +\sigma^2_{i}}   \right ) - \ln \left ( {\sigma^2_{v_1}{\left \| {\bf{v}}^{\rm{H}}{\rm{diag}}\left({\bf{h}}_{{\rm{R}}i}^{\rm{H}} \right) \right \|}^{2}_{2} +\sigma^2_{i}} \right ).
		\end{split}
	\end{equation}
\end{figure*}
We then apply the SDR method, let ${\bar{\bf{v}}} = \left[{\bf{v}}^{\rm{H}}, 1 \right]^{\rm{H}}  $, ${\bar{\bf{V}}} = {\bar{\bf{v}}}{\bar{\bf{v}}}^{\rm{H}}$, and the expression of $R_i$ in ({\ref{CICICI}}) can be further reformulated into a more tractable form as ${{R_{i}}}  = \ln\left( {\rm{tr}}\left ( {\bar{\bf{H}}}_{i_1}{\bar{\bf{V}}} \right )\right)  -\ln \left( {\rm{tr}}\left ( {\bar{\bf{H}}}_{i_2}{\bar{\bf{V}}} \right )\right) $, where ${\bar{\bf{H}}}_{i_1}$ is given by (\ref{asdefc}) at the bottom of this page, and
\begin{figure*}[hb]
\vspace{-0.5cm}
{\noindent}	 \rule[5pt]{18cm}{0.1em}\\
\vspace{-0.2cm}
\begin{equation}\label{asdefc}
	{\bar{\bf{H}}}_{i_1} = \begin{bmatrix}
		{\rm{diag}}\left ( {\bf{h}}_{{\rm{R}}i}^{\rm{H}} \right ) {\bf{H}}_{\rm{BR}}{\bf{w}}_{\rm{c}}{\bf{w}}_{\rm{c}}^{\rm{H}}{\bf{H}}_{\rm{BR}}^{\rm{H}}{\rm{diag}}\left ( {\bf{h}}_{{\rm{R}}i}^{\rm{H}} \right )^{\rm{H}} +{\bf{H}}_{i{\rm{N}}} &{\rm{diag}}\left ( {\bf{h}}_{{\rm{R}}i}^{\rm{H}} \right ) {\bf{H}}_{\rm{BR}}{\bf{w}}_{\rm{c}}{\bf{w}}_{\rm{c}}^{\rm{H}}{\bf{h}}_{{\rm{B}}i} \\
		{\bf{h}}_{{\rm{B}}i}^{\rm{H}}{\bf{w}}_{\rm{c}}{\bf{w}}_{\rm{c}}^{\rm{H}}{\bf{H}}_{\rm{BR}}^{\rm{H}}{\rm{diag}}\left ( {\bf{h}}_{{\rm{R}}i}^{\rm{H}} \right )^{\rm{H}} & {\sigma}_{i}^{2} + {\bf{w}}_{\rm{c}}^{\rm{H}}{\bf{h}}_{{\rm{B}}i}{\bf{h}}_{{\rm{B}}i}^{\rm{H}}{\bf{w}}_{\rm{c}}
	\end{bmatrix}.
\end{equation}
\end{figure*}
\begin{equation}\nonumber
	{\bar{\bf{H}}}_{i_2} = \begin{bmatrix}
		{\bf{H}}_{i\rm{N}} & {\bf{0}}_{N \times 1} \\
		{\bf{0}}_{1 \times N} & 0
	\end{bmatrix}, 
\end{equation}
where ${\bf{H}}_{i\rm{N}} = {\sigma}_{v_1}^{2}{\rm{diag}}\left ( {\bf{h}}_{{\rm{R}}i}^{\rm{H}} \right )\left(  {\rm{diag}}\left ( {\bf{h}}_{{\rm{R}}i}^{\rm{H}} \right )\right) ^{\rm{H}}$.

To tackle the quartic form of $\bf{\Phi}$ in constraints (\ref{ppppp3}b) and (\ref{ppppp3}c), we first denote ${\bf{V}}= {\bf{v}}{\bf{v}}^{\rm{H}}$ and ${\hat {\bf{v}}}= {\rm{vec}}\left( {\bf{V}} \right)$. By applying Lemma 1, the lower bound of  radar SINR in constraint ({\ref{ppppp3}}b) is given by
\begin{equation}\label{axxwer}
	\begin{split}
		\quad &{\rm{tr}}\left ( {\bf{A}}   {\bf{R}}  {\bf{A}}^{\rm{H}} {\bf{J}} ^{-1}  \right ) \\
		\ge & 2{\rm{Re}}\left({\rm{tr}}\left (  {\bf{A}}  {\bf{R}}   {\bf{A}}_{(l)}^{\rm{H}} {\bf{J}}_{(l)}^{-1} \right ) \right )- {\rm{tr}}\left ({\bf{J}}_{(l)}^{-1}{\bf{A}}_{(l)}{\bf{R}} {\bf{A}}_{(l)}^{\rm{H}}  {\bf{J}}_{(l)}^{-1} {\bf{J}} \right ),
	\end{split}
\end{equation}
where ${\bf{A}}_{(l)}, {\bf{J}}_{(l)}$ are the value of ${{\bf{A}}}, {{\bf{J}}}$ in the $l$-th iteration, respectively.

Then, by using the property ${\rm{tr}}\left ( {\bf{A}}^{\rm{H}}{\bf{B}} \right ) = \left( {\rm{vec}}\left ( {\bf{A}} \right )\right) ^{\rm{H}} {\rm{vec}}\left ( {\bf{B}} \right ) $ \cite{Matrix}, the first term on the right hand side of (\ref{axxwer}) can be reformulated as
\begin{equation}\label{abnjth}
\begin{split}
	&{\rm{tr}}\left (  {\bf{A}}  {\bf{R}}   {\bf{A}}_{(l)}^{\rm{H}} {\bf{J}}_{(l)}^{-1} \right )\\
	=&{\rm{tr}}\left ({\bf{H}}_{\rm{BR}} {\bf{R}}   {\bf{A}}_{(l)}^{\rm{H}} {\bf{J}}_{(l)}^{-1}{\bf{H}}_{\rm{BR}}^{\rm{H}}{\bf{\Phi}}^{\rm{H}}{\bf{G}}{\bf{\Phi}}\right )  \\
	=& \left( {\rm{vec}} \left ( {\bf{H}}_{\rm{BR}}{\bf{J}}_{(l)}^{-1}{\bf{A}}_{(l)}{\bf{R}}{\bf{H}}_{\rm{BR}}^{\rm{H}} \right ) \right) ^{\rm{H}} {\rm{vec}} \left (  {\bf{\Phi}}^{\rm{H}}{\bf{G}}{\bf{\Phi}} \right ).
\end{split}
\end{equation}
Furthermore, by using the property ${\rm{vec}} \left (  {\bf{A}}{\bf{B}}{\bf{C}} \right ) = \left ( {\bf{C}}^{\rm{T}} \otimes {\bf{A}}\right ) {\rm{vec}} \left (  {\bf{B}} \right )$\cite{Matrix}, we have
\begin{equation}\label{alopwe}
	\begin{split}
		 {\rm{vec}} \left (  {\bf{\Phi}}^{\rm{H}}{\bf{G}}{\bf{\Phi}} \right )
		=& \left ( {\bf{\Phi}} \otimes {\bf{\Phi}}^{\ast } \right ){\rm{vec}} \left (  {\bf{G}} \right )\\
		=&{\rm{diag}}\left( {\rm{vec}} \left (  {\bf{G}} \right )\right ){\hat{\bf{v}}},\\
	\end{split}
\end{equation}
where the second equation holds since ${\bf{\Phi}} \otimes {\bf{\Phi}}^{\ast } = {\rm{diag}}\left ( {\hat{\bf{v}}} \right ) $. By using (\ref{abnjth}) and (\ref{alopwe}), the first term on the right hand side of (\ref{axxwer}) can be reformulated as
\begin{equation}\label{solkjd}
2{\rm{Re}}\left({\rm{tr}}\left (  {\bf{A}}  {\bf{R}}   {\bf{A}}_{(l)}^{\rm{H}} {\bf{J}}_{(l)}^{-1} \right ) \right )=
2\rm{Re}\left( {\bf{p}}_1^{H}{\hat{\bf{v}}}\right),
\end{equation}
where
\begin{equation}\label{qwssdc}
{\bf{p}}_1^{\rm{H}}=\left( {\rm{vec}} \left ( {\bf{H}}_{\rm{BR}}{\bf{J}}_{(l)}^{-1}{\bf{A}}_{(l)}{\bf{R}}{\bf{H}}_{\rm{BR}}^{\rm{H}} \right ) \right) ^{\rm{H}}{\rm{diag}}\left( {\rm{vec}} \left (  {\bf{G}} \right )\right).
\end{equation}

By defining the constant matrix ${\bf{E}} = {\bf{J}}_{(l)}^{-1} {\bf{A}}_{(l)} {\bf{R}}{\bf{A}}_{(l)}^{\rm{H}}{\bf{J}}_{(l)}^{-1} \succeq {\bf{0}}$ in the $l$-th iteration, the second term on the right hand side of (\ref{axxwer}) is transformed into
\begin{equation}\label{apsmel}
	\begin{split}
		&{\rm{tr}}\left ({\bf{J}}_{(l)}^{-1} {\bf{A}}_{(l)} {\bf{R}}{\bf{A}}_{(l)}^{\rm{H}}{\bf{J}}_{(l)}^{-1} {\bf{J}} \right )\\   
		=\ &\begin{matrix}\underbrace{{\rm{tr}}\left ( {\bf{E}} {\bf{H}}_{\rm{BR}}^{\rm{H}}{\bf{\Phi}}{\bf{\Xi }}_1{\bf{\Phi}}^{\rm{H}}{\bf{H}}_{\rm{BR}} \right )} 
			\\{\textrm{quadratic term}}
		\end{matrix}  \\
		+\ &\begin{matrix}\underbrace{2{\sigma_{v_1}^{2}}{\rm{Re}}\left ( {\rm{tr}}\left ( {\bf{E}} {\bf{H}}_{\rm{BR}}^{\rm{H}}{\bf{\Phi}}^{\rm{H}}{\bf{G}}{\bf{\Phi}}{\bf{\Phi}}^{\rm{H}}{\bf{H}}_{\rm{BR}} \right ) \right )} 
			\\{\textrm{cubic term}}
		\end{matrix}\\
		+\ &\begin{matrix}\underbrace{{\sigma_{v_1}^{2}}{\rm{tr}}\left ( {\bf{E}} {\bf{H}}_{\rm{BR}}^{\rm{H}}{\bf{\Phi}}^{\rm{H}}{\bf{G}}{\bf{\Phi}}{\bf{\Phi}}^{\rm{H}}{\bf{G}}^{\rm{H}}{\bf{\Phi}}{\bf{H}}_{\rm{BR}} \right )}
			\\{\textrm{quartic term}}
		\end{matrix}\\
		+\ &{\sigma_{\rm{R}}^{2}}{\rm{tr}}\left ({\bf{E}}  \right ),
	\end{split}
\end{equation}
where ${\bf{\Xi}}_1 = \left({\sigma}_{v_1}^2+{\sigma}_{v_2}^2\right){\bf{I}}_N + {\rho}^2{\bf{H}}_{\rm{BR}}{\bf{R}}{\bf{H}}_{\rm{BR}}^{\rm{H}}$. It is worth noting that  (\ref{apsmel}) is a quartic expression with quadratic, cubic and quartic terms with respect to ${\bf{\Phi}}$. To tackle the high-order form of (\ref{apsmel}), we construct a lower bound of ${{\xi}}_{\rm{R}}$ with a more tractable low-order form. According to the property ${\rm{tr}}\left ( {\bf{ABCD}} \right )  = \left (  {\rm{vec}}\left ( {\bf{D}}^{\rm{T}} \right ) \right ) ^{\rm{T}}  \left ({\bf{C}}^{\rm{T}} \otimes {\bf{A}} \right )  {\rm{vec}}\left ( {\bf{B}} \right )$\cite{Matrix}, the quartic term on the left hand side of (\ref{apsmel}) can be transformed into a quadratic expression of $\hat{\bf{v}}$.
\begin{equation}\label{ioskld}
	\begin{aligned}
		&{\sigma_{v_1}^{2}}{\rm{tr}}\left ( {\bf{E}} {\bf{H}}_{\rm{BR}}^{\rm{H}}{\bf{\Phi}}^{\rm{H}}{\bf{G}}{\bf{\Phi}}{\bf{\Phi}}^{\rm{H}}{\bf{G}}^{\rm{H}}{\bf{\Phi}}{\bf{H}}_{\rm{BR}}\right ) \\
		=&{\sigma_{v_1}^{2}}{\rm{tr}}\left ( {\bf{H}}_{\rm{BR}}{\bf{E}} {\bf{H}}_{\rm{BR}}^{\rm{H}}{\bf{\Phi}}^{\rm{H}}{\bf{G}}{\bf{\Phi}}{\bf{I}}_{N}{\bf{\Phi}}^{\rm{H}}{\bf{G}}^{\rm{H}}{\bf{\Phi}}  \right)\\
		=&{\sigma_{v_1}^{2}}\left({\rm{vec}}\left ({\bf{\Phi}}^{\rm{H}}{\bf{G}}{\bf{\Phi}} \right)\right)^{\rm{H}}   \left ( {\bf{I}}_{N} \otimes {\bf{H}}_{\rm{BR}}{\bf{E}} {\bf{H}}_{\rm{BR}}^{\rm{H}}  \right )     {\rm{vec}}\left ( {\bf{\Phi}}^{\rm{H}}{\bf{G}}{\bf{\Phi}}  \right)\\
		=&{\hat{\bf{v}}}^{\rm{H}}{\bf{Q}}_{1}{\hat{\bf{v}}},
	\end{aligned}
\end{equation}
where matrix
$
{\bf{Q}}_{1} =  {\sigma_{v_1}^{2}}{\rm{diag}} \left ( {\rm{vec}} \left ( {\bf{G}} \right ) \right )^{\rm{H}}   \left ( {\bf{I}}_{N} \otimes {\bf{H}}_{\rm{BR}}{\bf{E}} {\bf{H}}_{\rm{BR}}^{\rm{H}}  \right )$ ${\rm{diag}} \left ( {\rm{vec}} \left ( {\bf{G}} \right ) \right ).
$
Then, similar to obtaining (\ref{solkjd}), the quadratic term on the right hand side of (\ref{apsmel}) can be transformed into
\begin{equation}\label{lsmdwe}
\begin{split}
	&{\rm{tr}}\left ( {\bf{E}} {\bf{H}}_{\rm{BR}}^{\rm{H}}{\bf{\Phi}}{\bf{\Xi }}_1{\bf{\Phi}}^{\rm{H}}{\bf{H}}_{\rm{BR}} \right )\\
	=&\left( {\rm{vec}}\left ( {\bf{\Xi}}_1 \right )\right) ^{\rm{H}}  {\rm{diag}}\left({\rm{vec}}\left ( {\bf{H}}_{\rm{BR}}{\bf{E}}{\bf{H}}_{\rm{BR}}^{\rm{H}} \right ) \right )   {\hat{\bf{v}}}\\
	=&{\bf{p}}_{2,1}^{\rm{H}}{\hat{\bf{v}}},
\end{split} 
\end{equation}
where 
\begin{equation}
{\bf{p}}_{2,1}^{\rm{H}} = \left( {\rm{vec}}\left ( {\bf{\Xi}}_1 \right )\right) ^{\rm{H}}  {\rm{diag}}\left({\rm{vec}}\left ( {\bf{H}}_{\rm{BR}}{\bf{E}}{\bf{H}}_{\rm{BR}}^{\rm{H}} \right ) \right ).
\end{equation}
Finally, we can obtain a tractable upper bound of the cubic term on the right hand side of (\ref{apsmel}) by using the following lemma.

\textbf{\emph{Lemma 3}}: Given ${\bf{K}}_{(l)}, {\bf{L}}_{(l)}$ as the value of ${\bf{K}}, {\bf{L}}$ in the $l$-th iteration, we have
\begin{equation}\nonumber
\begin{split}
&\ 2{\rm{Re}}\left ( {\rm{tr}}\left ( {\bf{K}}{\bf{L}}^{\rm{H}} \right )  \right )\\ \le& \ \frac{\left \| {\bf{L}}_{(l)}  \right \|_{\rm{F}} }{\left \|  {\bf{K}}_{(l)}  \right \|_{\rm{F}}} {\rm{tr}}\left ( {\bf{KK}}^{\rm{H}} \right )+ \frac{\left \| {\bf{K}}_{(l)}  \right \|_{\rm{F}} }{\left \|  {\bf{L}}_{(l)}  \right \|_{\rm{F}}} {\rm{tr}}\left ( {\bf{LL}}^{\rm{H}} \right ).
\end{split}
\end{equation}

\textbf{\emph{Proof}}: The above inequality can be obtained through the expansion of $\left \| \sqrt{\frac{\left \| {\bf{L}}_{(l)}  \right \|_{\rm{F}} }{\left \|  {\bf{K}}_{(l)}  \right \|_{\rm{F}}}}   {\bf{K}} -  \sqrt{\frac{\left \| {\bf{K}}_{(l)}  \right \|_{\rm{F}} }{\left \|  {\bf{L}}_{(l)}  \right \|_{\rm{F}}}}   {\bf{L}} \right \|_{\rm{F}}^{2} \ge 0$.$\hfill\blacksquare$

Based on Lemma 3, we have
\begin{equation}\label{scxbds}
	\begin{split}
		&2{\sigma_{v_1}^{2}}{\rm{Re}}\left ( {\rm{tr}}\left ( {\bf{E}} {\bf{H}}_{\rm{BR}}^{\rm{H}}{\bf{\Phi}}^{\rm{H}}{\bf{G}}{\bf{\Phi}}{\bf{\Phi}}^{\rm{H}}{\bf{H}}_{\rm{BR}} \right )  \right )\\
		\le&{\sigma_{v_1}^{2}}\left ( \beta^2\left \| {\bf{\Phi}}^{\rm{H}}{\bf{H}}_{\rm{BR}}{\bf{E}}{\bf{H}}_{\rm{BR}}^{\rm{H}} \right \|_{\rm{F}}^{2}  + \frac{1}{\beta^2}\left \| {\bf{\Phi}}^{\rm{H}}{\bf{G}}{\bf{\Phi}} \right \|_{\rm{F}}^{2}  \right ) \\
		=&{\bf{p}}_{2,2}^{\rm{H}}{\hat{\bf{v}}}+{\hat{\bf{v}}}^{\rm{H}}{\bf{Q}}_2{\hat{\bf{v}}},
	\end{split} 
\end{equation}
where ${\beta}^{2} = \left \| {\bf{\Phi}}_{(l)}^{\rm{H}}{\bf{G}}{\bf{\Phi}}_{(l)} \right \|_{\rm{F}}/\left \| {\bf{\Phi}}_{(l)}^{\rm{H}}{\bf{H}}_{\rm{BR}}{\bf{E}}{\bf{H}}_{\rm{BR}}^{\rm{H}} \right \|_{\rm{F}}$ and ${\bf{\Phi}}_{(l)}$ denotes the value of ${\bf{\Phi}}$ in the $l$-th iteration,  ${\bf{p}}_{2,2}^{\rm{H}} = {\sigma}_{v_1}^{2}{\beta}^{2}\left ( {\rm{vec}}\left ({\bf{H}}_{\rm{BR}}{\bf{E}}{\bf{H}}_{\rm{BR}}^{\rm{H}}{\bf{H}}_{\rm{BR}}{\bf{E}}{\bf{H}}_{\rm{BR}}^{\rm{H}}  \right )  \right )^{\rm{H}}  {\rm{diag}}\left ( {\rm{vec}}\left ( {\bf{I}}_{N} \right )  \right ), $ and
\begin{equation}\label{qwllsp}
	{\bf{Q}}_2 =\frac{{{\sigma}_{v_1}^{2}}}{{{\beta}^{2}}} \left ( {\rm{diag}}\left ({\rm{vec}}\left ( \bf{G} \right )   \right )  \right ) ^{\rm{H}}\left ( {\rm{diag}}\left ({\rm{vec}}\left ( \bf{G} \right )   \right )  \right ),
\end{equation} 
respectively. Substituting (\ref{solkjd}), (\ref{ioskld}), (\ref{lsmdwe}) and (\ref{scxbds}) into (\ref{axxwer}), a more tractable lower bound of ${\xi}_{\rm{R}}$ as a quadratic function of ${\hat{\bf{v}}}$ is given as
\begin{equation}\label{aswwer}
		{\xi}_{\rm{R}}\ge{\rm{2Re}}\left ( {\bf{p}}_{1}^{\rm{H}} {\hat{\bf{v}}} \right ) -{\bf{p}}_{2}^{\rm{H}}{\hat{\bf{v}}} -{\hat{\bf{v}}}^{\rm{H}}{\bf{Q}}_{1}{\hat{\bf{v}}}-{\hat{\bf{v}}}^{\rm{H}}{\bf{Q}}_2{\hat{\bf{v}}}-\alpha_2,
\end{equation}
where ${\bf{p}}_{2} \triangleq {\bf{p}}_{2,1}+{\bf{p}}_{2,2}$, and $\alpha_2 \triangleq {\sigma_{\rm{R}}^{2}}{\rm{tr}}\left ({\bf{E}}  \right )$. Since equation ${\hat{\bf{v}}} = {\rm{vec}}\left ( {\bf{V}} \right ) $ holds, we can rewrite the right hand side of (\ref{aswwer}) as a function of  ${{\bf{V}}}$ as
\begin{equation}\label{asdfty}
	\begin{split}
		&{\rm{2Re}}\left ( {\bf{p}}_{1}^{\rm{H}} {\hat{\bf{v}}} \right ) -{\bf{p}}_{2}^{\rm{H}}{\hat{\bf{v}}} -{\hat{\bf{v}}}^{\rm{H}}{\bf{Q}}_{1}{\hat{\bf{v}}}-{\hat{\bf{v}}}^{\rm{H}}{\bf{Q}}_2{\hat{\bf{v}}}-\alpha_2\\
		=& -{\rm{tr}}\left ( {\bf{V}}{\bf{M}}_{1,1}{\bf{V}}^{\rm{H}} \right )-{\rm{tr}}\left ( {\bf{V}}{\bf{M}}_{1,2}{\bf{V}}^{\rm{H}} \right )\\
		& + {\rm{tr}}\left ( {\bf{N}}_{1,1}{\bf{V}} \right )+ {\rm{tr}}\left ( {\bf{V}}^{\rm{H}}{\bf{N}}_{1,2} \right ) - \alpha_2\\
		=&-{\rm{tr}}\left ( {\bf{V}}{\bf{M}}_{1}{\bf{V}}^{\rm{H}}\right ) + {\rm{tr}}\left ( {\bf{N}}_{1}{\bf{V}} \right ) - \alpha_2,
	\end{split}
\end{equation}
where
\begin{equation}\label{aserjk}
	{\bf{M}}_{1,1}=
	 {\bf{Q}}_{1\left[ 1:N,1:N\right]},
	{\bf{M}}_{1,2}=
	 {\bf{Q}}_{2\left[ 1:N,1:N\right]} ,
\end{equation}
\begin{equation}\label{axcrrt}
	{\bf{N}}_{1,1}^{\rm{H}} = \Sigma \left ( {\bf{p}}_{1}-{\bf{p}}_{2}\right ) ,
	{\bf{N}}_{1,2} = \Sigma \left ( {\bf{p}}_{1} \right ) ,
\end{equation}
\begin{equation}\label{aofrrt}
	{\bf{M}}_{1} = {\bf{M}}_{1,1} + {\bf{M}}_{1,2},{\bf{N}}_{1} = {\bf{N}}_{1,1} + {\bf{N}}_{1,2}.
\end{equation}

The proof of (\ref{aserjk}) and the positive semi-definiteness of ${\bf{M}}_1$ can be found in Appendix B. We then transform the expression of the transmit power of the active RIS into the quadratic form of ${\bf{V}}$. In terms of the left hand side of constraint (\ref{ppppp3}c), we have
\begin{equation}\label{aswwqq}
P_{\rm{{A_1}}}+{P_{\rm{A_2}}}
={\rm{tr}}\left ( {\bf{\Xi}}_2{\bf{\Phi}}^{\rm{H}} {\bf{\Phi}}\right )
+ {\rm{tr}}\left ( {\bf{\Xi}}_3{\bf{\Phi}}^{\rm{H}}{\bf{G}}^{\rm{H}}{\bf{\Phi}}{\bf{\Phi}}^{\rm{H}}{\bf{G}}{\bf{\Phi}} \right ),
\end{equation}
where ${\bf{\Xi}}_2 = {\bf{H}}_{\rm{BR}}{\bf{R}}{\bf{H}}_{\rm{BR}}^{\rm{H}}+\left ( {\sigma}_{v_1}^2 + {\sigma}_{v_2}^2 \right ){\bf{I}}_N $, ${\bf{\Xi}}_3 = {\bf{H}}_{\rm{BR}}{\bf{R}}{\bf{H}}_{\rm{BR}}^{\rm{H}}+ {\sigma}_{v_1}^2 {\bf{I}}_N $, respectively. Then the right hand side of (\ref{aswwqq}) can be reformulated as a function of ${\hat{\bf{v}}}$ as
\begin{equation}\label{opskem}
	\begin{split}
		&\ {\rm{tr}}\left ( {\bf{\Xi}}_2{\bf{\Phi}}^{\rm{H}} {\bf{\Phi}}\right )
		+ {\rm{tr}}\left ( {\bf{\Xi}}_3{\bf{\Phi}}^{\rm{H}}{\bf{G}}^{\rm{H}}{\bf{\Phi}}{\bf{\Phi}}^{\rm{H}}{\bf{G}}{\bf{\Phi}} \right )\\
	=&\ {\hat{\bf{v}}}^{\rm{H}}{\bf{Q}}_{3}{\hat{\bf{v}}}+{\bf{p}}_{3}^{\rm{H}}{\hat{\bf{v}}},
	\end{split}
\end{equation}
where ${\bf{Q}}_3$ and ${\bf{p}}_3$ are respectively given as
\begin{align}
	{\bf{Q}}_{3} &=  {\rm{diag}} \left ( {\rm{vec}} \left ( {\bf{G}} \right ) \right )^{\rm{H}}   \left ( {\bf{I}}_{N} \otimes {\bf{\Xi}}_3  \right ){\rm{diag}} \left ( {\rm{vec}} \left ( {\bf{G}} \right ) \right ),\label{asdenm}\\
	{\bf{p}}_{3}^{\rm{H}} &=\left(  {\rm{vec}}\left ( {\bf{\Xi}}_2 \right )\right)^{\rm{H}}{\rm{diag}}\left ( {\rm{vec}}\left ( {\bf{I}}_{N} \right )  \right ).\label{apswle}
\end{align}

In addition, the expression of (\ref{opskem}) as a function of ${\bf{V}}$ is given as
\begin{equation}\label{axeerr}
		{\hat{\bf{v}}}^{\rm{H}}{\bf{Q}}_{3}{\hat{\bf{v}}}+{\bf{p}}_{3}^{\rm{H}}{\hat{\bf{v}}}
		={\rm{tr}}\left ( {\bf{V}}{\bf{M}}_{2}{\bf{V}}^{\rm{H}} \right ) + {\rm{tr}}\left ( {\bf{N}}_{2}{\bf{V}} \right ),
\end{equation}
where
\begin{equation}\label{scrtgh}
	{\bf{M}}_{2}=
	{\bf{Q}}_{{3}{\left[ 1:N,1:N\right]}},
	{\bf{N}}_{2}^{\rm{H}} = \Sigma \left ( {\bf{p}}_{3}\right ).
\end{equation}

Note that matrix ${\bf{M}}_{2}$ is also positive semi-definite, and the detailed proof can be found in Appendix B. 
Finally, the constraint (\ref{ppppp3}d) can be reformulated as ${\bar{\bf{V}}}_{\left [ i,i  \right ] }   \le \eta_i^{2} ,i = 1,2,\cdots, N $ , ${\bar{\bf{V}}}_{\left [ N+1,N+1  \right ]  }   = 1 $. With the rank-1 constraint relaxed, Problem (\ref{ppppp3}) can be reformulated as
\begin{subequations}\label{ppppp6}
	\begin{align}
		\mathop {\max }\limits_{{{\bar{\bf{V}} }} } \quad &R\left ( \bar{\bf{V}} \right ) \\
		\qquad\ \textrm{s.t.}\quad
		&{\left\| {\bf{V}}{\bf{L}}_{1} \right\|_{\rm{F}}^{2}}- {\rm{tr}}\left ( {\bf{N}}_{1}{\bf{V}} \right )+e_2   \le  0,\\
		&{\left\| {\bf{V}}{\bf{L}}_{2} \right\|_{\rm{F}}^{2}}+ {\rm{tr}}\left ( {\bf{N}}_{2}{\bf{V}} \right )  \le P_{\rm{RIS}},\\
		&{\bar{\bf{V}}}_{\left [ i,i  \right ]}    \le \eta_i^{2} ,i = 1,2,\cdots, N, \\
		&{\bar{\bf{V}}}_{\left [ N+1,N+1  \right ]}     = 1,\\
		&{\bar{\bf{V}}}\succeq  {\bf{0}},
	\end{align}
\end{subequations}
where $e_2 = \alpha_2 + \gamma_{\rm{r}}$, ${\bf{L}}_i $ are the Cholesky decomposition of positive semi-definite matrix ${\bf{M}}_i$, namely ${\bf{M}}_i  = {\bf{L}}_i{\bf{L}}_{i}^{\rm{H}}, i = 1,2$, and the detailed expression of $R\left ( \bar{\bf{V}} \right )$ is shown in (\ref{ascfge}) at the bottom of the next page. Since ${\bf{V}} = {\bar{\bf{V}}}_{\left[ 1:N,1:N\right]} $ is an affine function of optimization variable ${\bar{\bf{V}}}$, it has no influence on the curvature of the constraints.

\begin{figure*}[hb]
	\vspace{-0.3cm}
	{\noindent}	 \rule[5pt]{18cm}{0.1em}\\
	\vspace{-0.3cm}
	\begin{equation}\label{ascfge}
			R\left ( \bar{\bf{V}} \right )  =\ln\left( {\rm{tr}}\left ( {\bar{\bf{H}}}_{\rm{U1}}{\bar{\bf{V}}}\right ) \right ) -\ln \left({\rm{tr}}\left ( {\bar{\bf{H}}}_{\rm{U2}}{\bar{\bf{V}}}\right ) \right )-\ln \left( {\rm{tr}}\left ( {\bar{\bf{H}}}_{\rm{E1}}{\bar{\bf{V}}} \right ) \right )+\ln \left( {\rm{tr}}\left ( {\bar{\bf{H}}}_{\rm{E2}}{\bar{\bf{V}}}\right ) \right ).
	\end{equation}
\end{figure*}

Similar to ({\ref{ppppp4}}a), we adopt the MM algorithm to tackle the non-convexity of the objective function. By using the first-order Taylor approximation at the given point ${\bar{\bf{V}}}_{\left(l \right) }$, the objective function can be transformed into (\ref{aoppsl}) at the bottom of this page.
\begin{figure*}[hb]
\vspace{-0.7cm}
\begin{equation}\label{aoppsl}
\begin{split}
	{\tilde{R}}\left ( {\bar{\bf{V}}} | {\bar{\bf{V}}}_{\left ( l \right )}  \right )
	 = & \ln \left ({\rm{tr}} \left ( {\bar{\bf{H}}}_{\rm{U1}}  {\bar{\bf{V}}}\right ) \right)-\ln \left ({\rm{tr}} \left ( {\bar{\bf{H}}}_{\rm{U2}}{\bar{\bf{V}}}_{\left ( l \right )} \right )\right)-{\rm{tr}}\left (  \frac{{\bar{\bf{H}}}_{\rm{U2}}}{{\rm{tr}}\left ( {\bar{\bf{H}}}_{\rm{U2}}{{\bar{\bf{V}}}_{\left ( l \right )} }\right )}  \left ( {\bar{\bf{V}}}-{{\bar{\bf{V}}}_{\left ( l \right )} } \right )   \right )\\
	 + & \ln \left ({\rm{tr}} \left ( {\bar{\bf{H}}}_{\rm{E2}}  {\bar{\bf{V}}}\right ) \right)-\ln \left ({\rm{tr}} \left ( {\bar{\bf{H}}}_{\rm{E1}}{\bar{\bf{V}}}_{\left ( l \right )} \right )\right)-{\rm{tr}}\left (  \frac{{\bar{\bf{H}}}_{\rm{E1}}}{{\rm{tr}}\left ( {\bar{\bf{H}}}_{\rm{E1}}{{\bar{\bf{V}}}_{\left ( l \right )} }\right )}  \left ( {\bar{\bf{V}}}-{{\bar{\bf{V}}}_{\left ( l \right )} } \right )   \right ) .
\end{split}
\end{equation}
\end{figure*}
By dropping the constant terms of ${\tilde{R}}\left ( {\bar{\bf{V}}} | {\bar{\bf{V}}}_{\left ( l \right )}   \right ) $, Problem (\ref{ppppp6}) is reformulated as
\begin{subequations}\label{ppppp7}
	\begin{align}
		\mathop {\max }\limits_{{{\bar{\bf{V}} }} } \quad & \ln \left ({\rm{tr}} \left ( {\bar{\bf{H}}}_{\rm{U_1}}  {\bar{\bf{V}}}\right ) \right)-{\rm{tr}}\left (  \frac{{\bar{\bf{H}}}_{\rm{U_2}}}{{\rm{tr}}\left ( {\bar{\bf{H}}}_{\rm{U_2}}{{\bar{\bf{V}}}_{\left ( l \right )}  }\right )}   {\bar{\bf{V}}}    \right )
		\\ 
		+	& \ln \left ({\rm{tr}} \left ( {\bar{\bf{H}}}_{\rm{E_2}}  {\bar{\bf{V}}}\right ) \right)-{\rm{tr}}\left (  \frac{{\bar{\bf{H}}}_{\rm{E_1}}}{{\rm{tr}}\left ( {\bar{\bf{H}}}_{\rm{E_1}}{{\bar{\bf{V}}}_{\left ( l \right )}  }\right )}  {\bar{\bf{V}}}   \right )\nonumber \\ 
		\qquad\ \textrm{s.t.}\quad
		& \textrm{(\ref{ppppp6}b)}\sim \textrm{(\ref{ppppp6}f)}, \nonumber
	\end{align}
\end{subequations}
which is a concave maximization problem and can be solved via CVX solver. The convergence of optimizing ${\bf{\Phi}}$ is also non-decreasing and we can iteratively obtain the optimal ${\bar{\bf{V}}}$. Similarly, the rank-1 solution of ${\bar{\bf{V}}}$ as ${\bar{\bf{V}}}^{\rm{opt}}$ can be recovered by applying the eigenvalue decomposition and Gaussian randomization methods, and the optimal solution ${\bf{\Phi}}^{\rm{opt}}$ can be obtained through ${\bar{\bf{V}}}^{\rm{opt}}$ as ${\bf{\Phi}}^{\rm{opt}} = {\rm{diag}}\left(\left( {\bm{\nu}}_{\rm{max}}\left ( {\bar{\bf{V}}}^{\rm{opt}} \right )_{\left[ 1:N\right]} \right) \sqrt{\lambda_{\rm{max}}  \left ( {\bar{\bf{V}}}^{\rm{opt}} \right )}\right )$. We summarize the procedures of the overall SDR-MM-based AO algorithm to optimize ${\bf{W}}$ and ${\bf{\Phi}}$ in Algorithm 1.
\begin{algorithm}
	\caption{SDR-MM-based AO algorithm.}
	\begin{algorithmic}[1]
		\STATE Initialize iteration number $t$ = 0, the maximum numbers of outer-layer iterations $t^{\rm{max}}$ and inner-layer iterations $\tau^{\rm{max}}_1, \tau^{\rm{max}}_2$. Initialize ${\bf{W}}^{[0]}$ and ${\bf{\Phi}}^{[0]}$.
		\STATE {\textbf{Repeat:}}
		\STATE \quad Let $k = 0$.
		\STATE \quad {\textbf{Repeat:}}
		\STATE \qquad Calculate ${\bf{W}}_{i(k+1)}$ according to (\ref{ppppp5}). 
		\STATE \qquad Recover the rank-1 solution by eigenvalue decomposition and Gaussian randomization.
		\STATE \qquad Let $k = k+1$.
		\STATE \quad {\textbf{Until}} {\textit{Convergence}} {\textbf{or}} $k = \tau^{\rm{max}}_1$.
		\STATE \quad Let ${\bf{W}}^{[t]} = {\bf{W}}_{(k+1)}$, and $l = $ 0.
		\STATE \quad {\textbf{Repeat:}}
		\STATE \qquad Calculate ${\bf{\Phi}}_{(l+1)}$ according to (\ref{ppppp7}).
		\STATE \qquad Recover the rank-1 solution by eigenvalue decomposition and Gaussian randomization.
		\STATE \qquad Let $l = l+1$.
		\STATE \quad {\textbf{Until}} {\textit{Convergence}} {\textbf{or}} $l = \tau^{\rm{max}}_2$.
		\STATE \quad Let ${\bf{\Phi}}^{[t]} = {\bf{\Phi}}_{(l+1)}$, and $t = t+1$.
		\STATE {\textbf{Until}} {\textit{Convergence}} {\textbf{or}} $t = t^{\rm{max}}$.
	\end{algorithmic}
\end{algorithm}

\subsection{Algorithm Analysis}
\subsubsection{Convergence Analysis}
According to the aforementioned analysis, the convergence of the MM algorithm is nondecreasing.  Since our variables ${\bf{W}}$ and ${\bf{\Phi}}$ are bounded by constraints, the convergence of the proposed SDR-MM-based AO algorithm is also non-decreasing and a sub-optimal point of our original problem can be obtained when the SDR-MM-based AO algorithm converges.

\subsubsection{Complexity Analysis}
The computational complexity of solving Problem (\ref{ppppp5}) and (\ref{ppppp7}) mainly lies in the interior point method, which is given by \cite{zhou2020framework} 
\begin{equation}\nonumber
	{\cal O}{\left ( \left ( \sum_{j=1}^{J}{k_j}+2m  \right )^{\frac{1}{2}}n \left (n^2+ \sum_{j=1}^{J}\left( {k_j^2}+{k_j^3}\right)+  n\sum_{i=1}^{m}{a_i^2}\right )  \right ) },
\end{equation}
where $n$ denotes the number of variables, $J$ denotes the number of linear matrix inequality (LMI) constraints, $k_j$ denotes the size of the $j$-th LMI constraint, $m$ denotes the number of second-order cone (SOC) constraints, and $a_i$ is the size of the $i$-th SOC constraint.

Problem (\ref{ppppp5}) contains $J_1 = \left( M+4\right) $ LMI constraints of size $k_1 = {\left(M+1 \right) }$ and the number of variables is $n_1 = {\left(M+1 \right) }^3$. Ignoring the constant value, the approximate computation complexity of Problem (\ref{ppppp5}) is given as $o_1 = {\cal O}{\left ( \left ( J_1 k_1  \right )^{1/2}n_1 \left (n_1^2+ n_1J_1 k_1^2 + J_1 k_1^3  \right )  \right ) } $. Similarly, the approximate computational complexity
of solving Problem (\ref{ppppp7}) with $J_2 = 2, k_2 = \left( N+1\right), n_2 = \left(N+1 \right)^2, m_2 = 2, a_2 = \left( N+1\right)^2   $ is given by $o_2 = {\cal O}{\left ( \left ( J_2 k_2+2m_2   \right )^{1/2}n_2 \left (n_2^2+ n_2J_2 k_2^2 + J_2 k_2^3 +n_2m_2 a_2 \right )  \right ) } $.
As a result, defining $t_{\rm{AO}}$ as the number of iterations of the AO algorithm, $t_1$ as the number of iterations of beamforming matrix optimization, and $t_2$ as the number of iterations of reflecting coefficient matrix optimization, the overall computational complexity of the proposed algorithm is given as ${t_{\rm{AO}}\left(t_1o_1+t_2o_2 \right) }$.

\section{Simulation Results}\label{simres}
In this section, simulation results are provided to illustrate the advantage of integrating an active RIS into the DFRC system for enhancing the dual-functional performance.

\subsection{Simulation Setup}\label{simset}
\subsubsection{Communication Channels}
According to the aforementioned system model, the location of the active RIS is designed carefully with few obstacles in the wireless environment, while the DFRC-BS may be located in a relatively crowded area. Thus, without loss of generality, channels ${\bf{H}}_{\rm{BR}}$, ${{\bf{h}}_{\rm{RU}}} $, and ${{\bf{h}}_{\rm{RE}}} $ are modeled as Rician fading, and channels ${{\bf{h}}_{\rm{BU}}} $ and ${{\bf{h}}_{\rm{BE}}} $ are modeled as Rayleigh fading, respectively.
The small-scale fading model of Rician channel is given as
\begin{equation}\label{axwwld}
	{\bar{\bf{H}}} = \sqrt{\frac{1}{\kappa +1} } {\bar{\bf{H}}_{\textrm{NLoS}}}+\sqrt{\frac{\kappa}{\kappa +1} } \bar{\bf{H}}_{\textrm{LoS}},
\end{equation}
where $\kappa$ denotes the Rician factor, $\bar{\bf{H}}_{\textrm{NLoS}}$ and $\bar{\bf{H}}_{\textrm{LoS}}$ denote the NLoS and LoS channel component between two devices, respectively. The NLoS component $\bar{\bf{H}}_{\textrm{NLoS}}$ follows Rayleigh fading and the LoS component $\bar{\bf{H}}_{\textrm{LoS}}$ is given as ${\bf{a}}_2\left ( {\theta_{2}} \right ){\bf{a}}_1^{\rm{H}}\left ( {\theta_{1}} \right )$, where
\begin{equation}\label{aaasas}
{\bf{a}}_1\left ( {\theta_{1}} \right )  = \left [ 1, e^{j2\pi\frac{ d_1}{\lambda} \rm{sin} {\theta_{1}}  } ,\dots, e^{j2\pi\frac{d_1}{\lambda} \left ( D_t-1 \right ) \rm{sin} {\theta_{1}} } \right ]^{\rm{H}},
\end{equation}
\begin{equation}\label{maasas}
{\bf{a}}_2\left ( {\theta_{2}} \right )  = \left [ 1, e^{j2\pi\frac{ d_2}{\lambda} \rm{sin} {\theta_{2}}  } ,\dots, e^{j2\pi\frac{d_2}{\lambda} \left ( D_r-1 \right ) \rm{sin} {\theta_{2}} } \right ]^{\rm{H}}.
\end{equation}
Parameters $D_r$ and $D_t$ denote the number of antennas/elements at the side of the receiver and transmitter, the angles ${\theta_{1}}$ and ${\theta_{2}}$ denote the AoD and the angle of arrival (AoA), and $d_1$, $d_2$ denote the intervals between adjacent antennas/elements, respectively.

The large-scale path loss in dB is given by
\begin{equation}\label{axwwOd}
	{\textrm{PL}}={\textrm{PL}}_0-10\alpha{\lg}\left(\frac{d}{d_0} \right) ,
\end{equation}
where ${\textrm{PL}}_0=$ -30 dB is the path loss at the reference distance $d_0={\textrm{1 m}}$, and $d$ is the link distance and $\alpha$ denotes the large-scale path loss exponent. 
\subsubsection{Sensing Response Channel}
To evaluate the fading nature of the target response channel, radar range equation is referenced to model the path loss coefficient $\gamma$ in matrix ${\bf{G}}$. By assuming that the target can be considered as a single scatter object, and the wireless sensing environment can be viewed as free space, the received radar echo power is given as
\begin{equation}\label{aWwsOi}
	P_{\rm{r}} = P_{\rm{t}}\frac{G^2\lambda^2\Delta}{\left ( 4\pi \right )^{3}R^{4} }, 
\end{equation}
where $P_t$ denotes the transmit power, $G$ denotes the array gain at the radar, $\lambda$ denotes the wavelength and $\Delta$ denotes the radar cross section (RCS) of the sensing target, respectively. Since the active RIS can also be viewed as a monostatic MIMO radar, the complex path loss coefficient can be modeled as
\begin{equation}\label{aWwwOi}
		\gamma = \sqrt{\frac{\lambda^2\Delta}{\left ( 4\pi \right )^{3}R^{4} }}, 
\end{equation}
where relative parameters are set as $f= $ 2.7 GHz and $\Delta={\textrm{1 m}}^2$\cite{richards2014fundamentals}. 

\subsubsection{Parameters Setup}
Unless otherwise stated, the simulation parameters are set as follows: Channel bandwidth of ${\rm{BW}} = $ 10 MHz, power density of thermal noise at the legitimate user, eavesdropper, radar receiver and active RIS of ${\sigma}^{2}  =  {\textrm{-174 dBm}}$\cite{zhi2022active}, number of transmit antennas of $M = $ 4, number of reflecting elements of active RIS of $N = $ 12, maximum power budget of the BS of $P_0 = $ 1 W, maximum power budget of the active RIS of $P_{\rm{RIS}} = $ 0.05 W, Rician factor of $\kappa = $ 3, SI coefficient of $\rho = $ 0.1, and the large-scale path loss exponent of Rayleigh and Rician channels are set to $\alpha_1 = $ 3.5 and $\alpha_2 = $ 2.2, respectively. For simplicity, the normalized intervals $d/\lambda$ are set to 0.5. Finally, the geometrical model is arranged as shown in Fig. 2.
\begin{figure}
	\centering
	\includegraphics[width=3.2in]{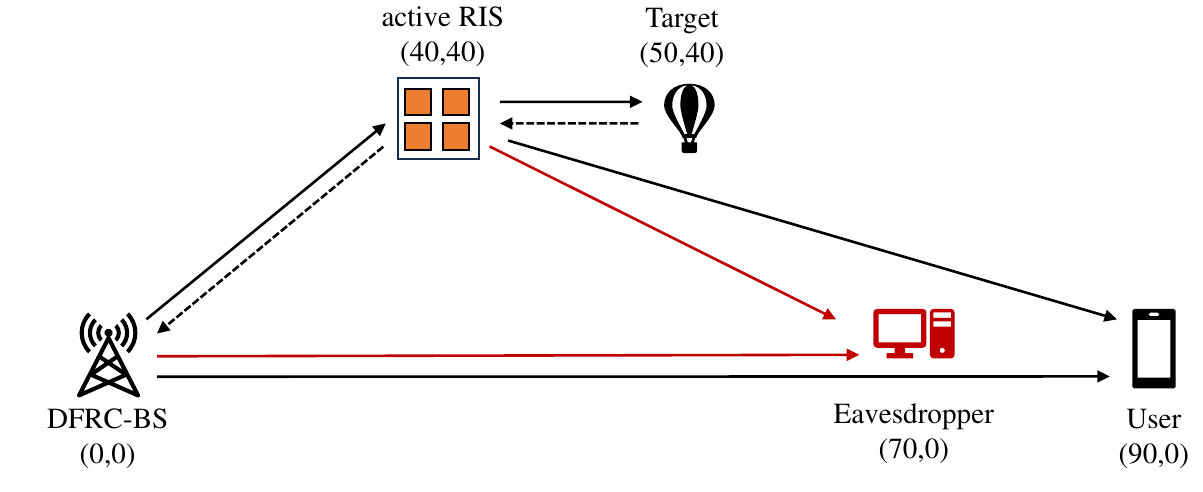}
	\caption{Simulation system setup.}
	\label{fig2}
\end{figure}

\subsection{Baseline Schemes}
We compare the performance of the proposed algorithms with the following baseline schemes.
\begin{itemize}
	\item[1)]
	\textbf{No RIS No Sensing}: To compare the SR gain obtained by the active and passive RISs, we implement a scheme with no RIS as a lower bound of SR performance with the sensing function ignored. 
\end{itemize}
\begin{itemize}
	\item[2)]
	\textbf{Passive RIS}: The passive RIS-related algorithm is almost the same as the original SDR-MM-based AO algorithm. Considering the property of the passive RIS, the terms generated by the amplified thermal noise and constraint (\ref{ppppp1}d) are dropped, and constraint (\ref{ppppp1}e) is reformulated as unit-modulus constraint.
\end{itemize}

For the fairness of our comparison, the total power consumption  is considered. The total power budgets of the three schemes are respectively given as
\begin{align}
		{Q^{\rm{act}}} &= {P^{\rm{act}}_{\rm{0}}}+{P_{\rm{RIS}}}+{N^{\rm{act}}}\left ( P_{\rm{SW}} + P_{\rm{DC}} \right )\\
		{Q^{\rm{pas}}} &= {P^{\rm{pas}}_{\rm{0}}}+{N^{\rm{pas}}} P_{\rm{SW}}\\
		{{Q^{\rm{NRNS}}}} &= {P^{\rm{NRNS}}_{\rm{0}}}
\end{align}
where ${Q^{\rm{act}}}, {Q^{\rm{pas}}}$ and ${{Q^{\rm{NRNS}}}}$ are the total power of the schemes ``Active RIS", ``Passive RIS" and ``No RIS No Sensing", and ${P_{0}^{\rm{act}}}, {P_{0}^{\rm{pas}}}$ and ${{P_{0}^{\rm{NRNS}}}}$ are the transmit power at the BS of the schemes ``Active RIS", ``Passive RIS" and ``No RIS No Sensing", respectively. The scalar $P_{\rm{SW}}$ = -5 dBm denotes the power consumption in phase control of each element of the RISs, and $P_{\rm{DC}}$ = -10 dBm denotes the direct-current (DC) power consumption of each active RIS element. After accounting for the power budget at the active RIS, the remaining power is allocated to the BS for the schemes ``Passive RIS" and ``No RIS No Sensing".

\subsection{Convergence Behavior of the Algorithm}
In this subsection, the convergence behavior of our algorithm is shown in Fig. 3. As we can see, the SDR-MM-based AO algorithm converges within 10 iterations to reach the target accuracy of ${10}^{-3}$ with different parameter settings, which indicates the good convergence behavior of our algorithm in different scenarios. In addition, it is shown that the secrecy performance can be enhanced by increasing the number of antennas, the number of RIS elements and the maximum amplitude gain of the active RIS.
\begin{figure}
	\centering
	\includegraphics[width=2.4in]{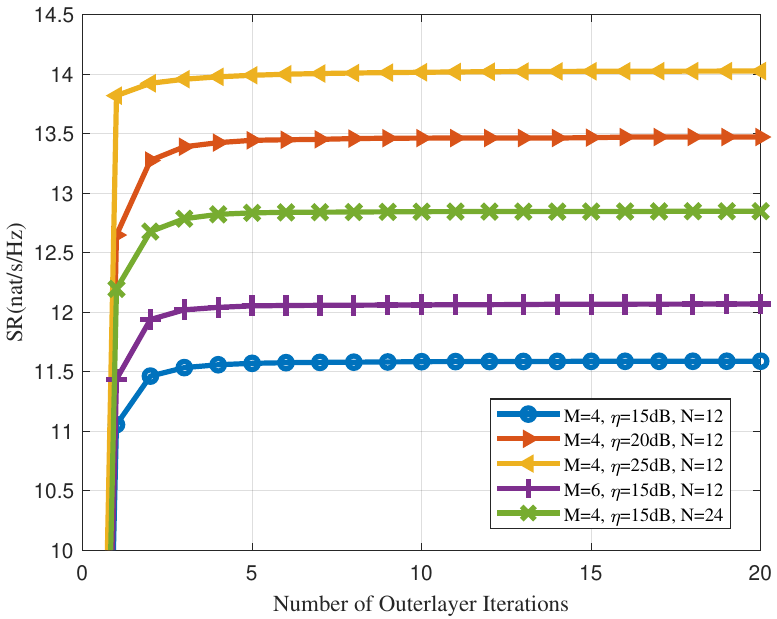}
	\caption{Convergence behavior of the proposed algorithm.}
	\label{fig3}
\end{figure}
\subsection{Relation between Communication and Sensing Function}
The sensing function and the trade-off between the two functions are investigated in this subsection. The cumulative distribution functions (CDF) of radar echo SINR are shown in Fig. 4, where the CDFs of the initial values are labeled as ``initialized" and those of the optimized values are labeled as ``optimized", respectively. The threshold ${\gamma_{\rm{{r}}}}$ is set as -80 dB.

As depicted in Fig. 4, the radar SINR of the active RIS-assisted system significantly exceeds that of the passive RIS-assisted one. Specifically, the mean SINR of the active RIS with $\eta = $ 15 dB exceeds that of the passive RIS by nearly 50 dB.
In addition, although the radar SINR is in constraint {(\ref{ppppp1}b)} instead of the objective function, the sensing performance of the active RIS-assisted system is still enhanced with the SDR-MM-based algorithm, while the sensing performance of the passive RIS-assisted system is actually degraded.
Therefore, we can conclude that the active RIS outperforms its passive counterpart in terms of sensing performance.

Furthermore, the relation between system SR and ${\gamma_{\rm{r}}}$ is shown in Fig. 5. The SR performance of the active RIS-assisted systems remains approximately constant with varying ${\gamma_{\rm{r}}}$. This also indicates that the radar sensing and secure communication function of our DFRC system can simultaneously reach their limit with well-designed parameters. In other words, the dual-function is jointly enhanced with no negative impact on each other in our proposed scheme.
\begin{figure}
	\centering
	\includegraphics[width=2.4in]{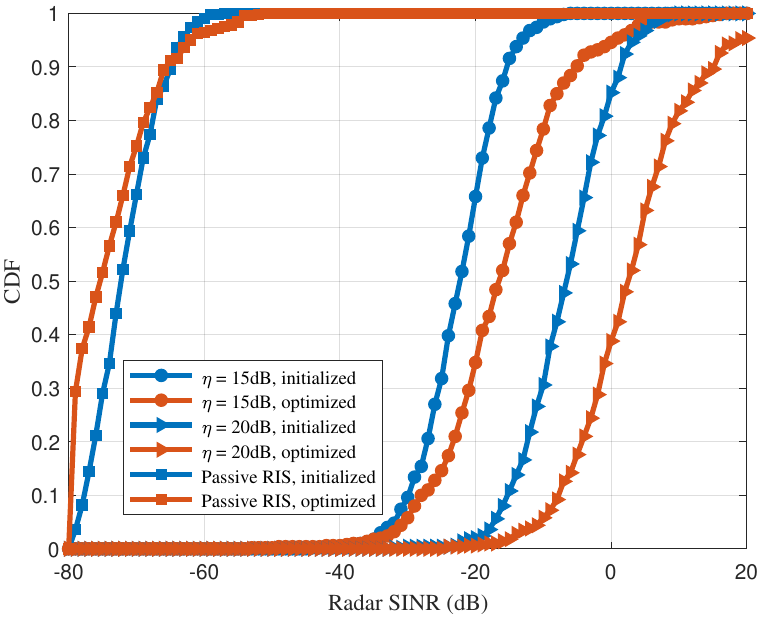}
	\caption{CDF of radar echo SINR.}
	\label{fig4}
\end{figure}
\begin{figure}
	\centering
	\includegraphics[width=2.4in]{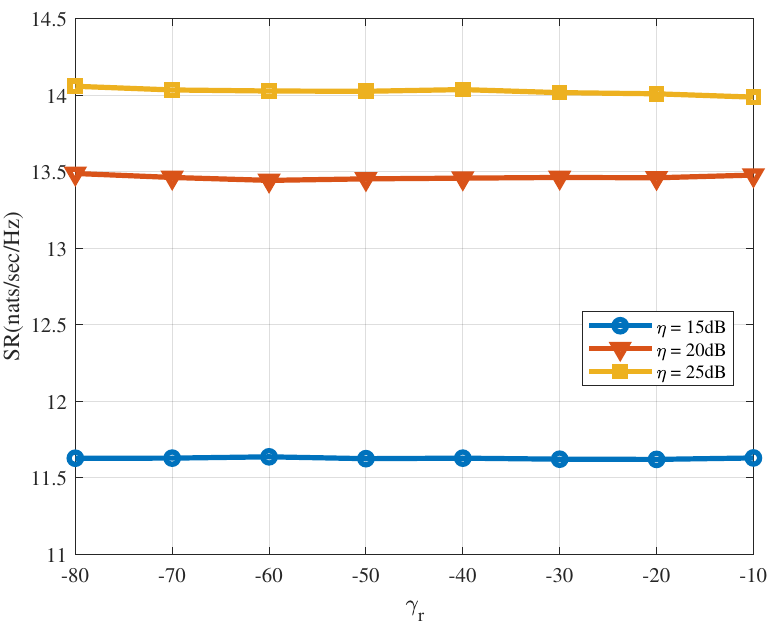}
	\caption{System SR versus ${\gamma_{\rm{r}}}$.}
	\label{fig5}
\end{figure}
\subsection{Impact of the Maximum Transmit Power}
The impact of the maximum transmit power at the BS and active RIS is illustrated in Fig. 6. It is obvious that there is an approximate logarithmic relationship between the SR and ${P_0}$.

As shown in Fig. 6, there is only a slight gain of the passive RIS-assisted system owing to the severe path loss of the RIS reflecting channel.
In contrast, the active RIS-assisted schemes achieve much better performance in terms of secrecy rate.
Specifically, when ${P_0} = $ 1 W, the deployment of the passive RIS obtains an SR gain of only 0.16 nat/s/Hz, in other words 1.68$\%$, while the active RIS with $\eta =$ 15/20 dB obtains a gain of 2.63/4.31 nat/s/Hz, in other words 27.2$\%$/44.6$\%$ respectively with an extra power consumption of 0.05 W.
This verifies the ability in combating the ``multiplicative fading" effect of the active RIS, and indicates the advantage in terms of secure wireless communication of deploying an active RIS over a passive RIS.

Furthermore, the transmit power budget at the active RIS also has an impact on the achievable SR. As illustrated in Fig. 6, the SR with ${P_{\rm{RIS}}} = $ 0.05 W and ${\eta = }$ 25 dB obtains a slight increase compared with the ${P_{\rm{RIS}}} = $ 0.05 W, ${\eta = }$ 20 dB scheme and finally reaches its limit, while the SR with ${P_{\rm{RIS}}} = $ 0.5 W, ${\eta = }$ 25 dB obviously exceeds the aforementioned counterparts. This is owing to the fact that the maximum transmit power at the active RIS is decisive for the maximum power received at the legitimate user. 
Therefore, we can conclude that  the maximum transmit power at the active RIS determines the achievable upper bound of SR in our proposed scheme.
\begin{figure}
	\centering
	\includegraphics[width=2.4in]{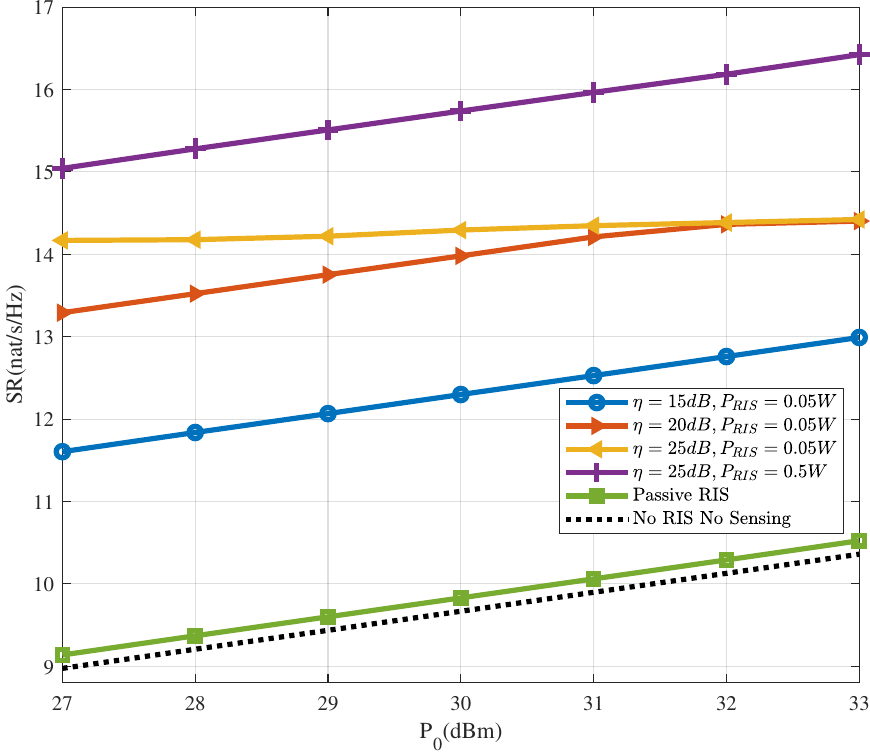}
	\caption{System SR versus the maximum transmit power.}
	\label{fig6}
\end{figure}
\subsection{Impact of the Number of RIS elements}
Fig. 7 illustrates the SR as a function of the number of reflecting elements. As expected, increasing the number of RIS elements enhances the secure communication performance. The diminishing return law and the impact of ${P_{\rm{RIS}}}$ are obvious in relevant schemes. Furthermore, the SR gain obtained by increasing the number of elements from 12 to 36 can exceed that of increasing the maximum amplitude gain by approximately 5 dB.
\begin{figure}
    \centering
    \includegraphics[width=2.4in]{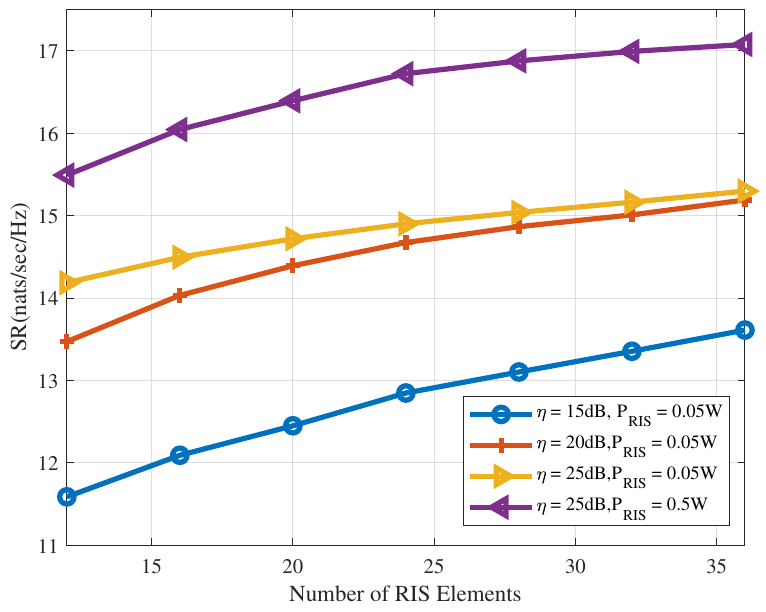}
    \caption{System SR versus the number of elements $N$.}
    \label{fig7}
\end{figure}
\subsection{Beampattern}
As is mentioned above, since the direct link between the DFRC-BS and the sensing target is obstructed, it is reasonable to view the active RIS as a monostatic MIMO Radar. Hence, we denote the beampattern at the active RIS as
\begin{equation}\label{bmpatn}
	\begin{split}
	P\left ( \theta \right ) =& {\mathbb{E}}\left\lbrace\left|  {\bf{a}}^{\rm{H}}\left ( \theta \right ){\bf{\Phi}}{{\bf{H}}_{\rm{BR}}}{\bf{x}}\right|^{2}\right\rbrace \\ =& {\bf{a}}^{\rm{H}}\left ( \theta \right ){\bf{\Phi}}{{\bf{H}}_{\rm{BR}}}{\bf{R}}{{\bf{H}}}_{\rm{BR}}^{\rm{H}}{\bf{\Phi}}^{\rm{H}}{\bf{a}}\left ( \theta \right ),
	\end{split}
\end{equation}
where $\theta$ denotes the angle of departure at the active RIS. We investigate the normalized beampattern gain with the legitimate user located at (90 m, 40 m, ${\theta_U} = 38.67^{\circ}$) and (103.78 m, 34.42 m, ${\theta_U} = 5^{\circ}$), which are shown in Fig. 8 and Fig. 9, respectively.

It is depicted that the center of the main lobe is located at the AoD of the legitimate user instead of the target, which is due to the fact that the objective function is not the radar SINR but the system SR.
In addition, the normalized beampattern shows a trough at the AoD of the eavesdropper and a side lobe at the AoD of the target (if not located in the main lobe), respectively.
Furthermore, it is worth noting that the
width of the main lobe is obviously reduced and the side lobes are effectively suppressed by increasing the number of RIS elements, while more antennas at the BS cannot lead to better directivity.
\begin{figure}
	\centering
	\includegraphics[width=2.4in]{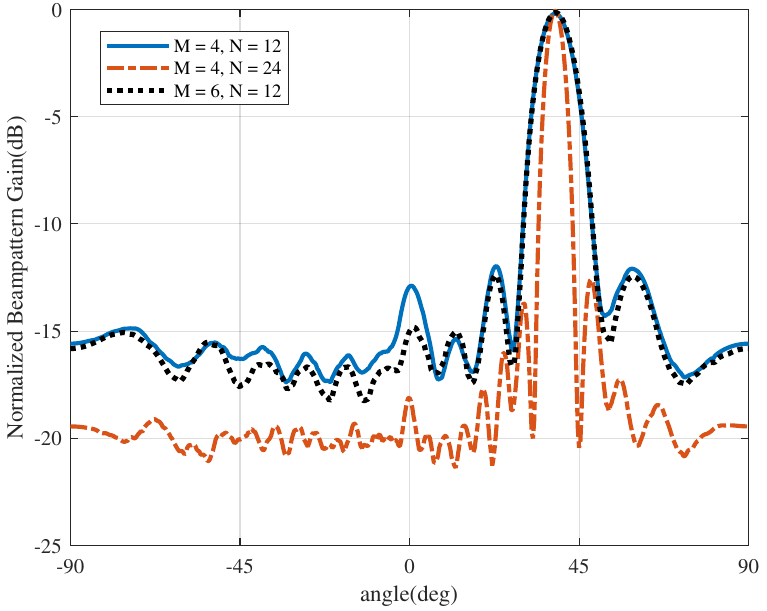}
	\caption{Normalized beampattern gain with $\theta_U = 38.67^{\circ}$.}
	\label{fig8}
\end{figure}
\begin{figure}
	\centering
	\includegraphics[width=2.4in]{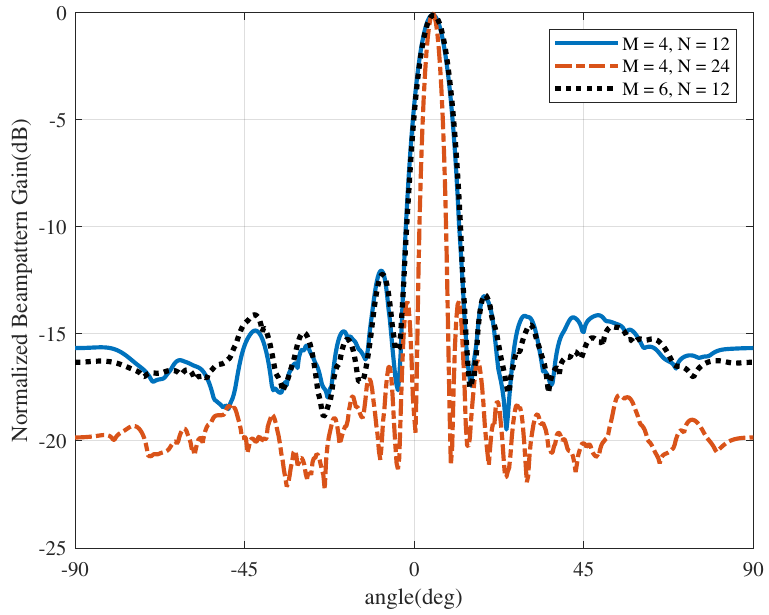}
	\caption{Normalized beampattern gain with $\theta_U = 5^{\circ}$.}
	\label{fig9}
\end{figure}

\section{Conclusions}\label{conclu}
In this paper, we studied an active RIS-assisted DFRC system with a potential eavesdropper and a four-hop sensing link. 
Generally, we aimed at solving the SR maximization problem by jointly optimizing the beamforming matrix at the DFRC-BS and the reflecting coefficient matrix at the active RIS, subject to the radar echo SINR constraint and the transmit power consumption constraints. An SDR-MM-based AO algorithm was proposed to tackle the complicated non-convex optimization problem. 
Our simulation results validated that the amplification function of the active RIS not only enhances the performance in physical layer security, but also observably overcomes the ``multiplicative fading" effect of the four-hop sensing channel. 
The active RIS prevails over traditional passive RIS in both secure communication as well as radar sensing with the same transmit power budget. 

\begin{appendices}
\section{Proof of Lemma 1}

Since function ${\rm{tr}}\left ( {{\bf{x}}^{\rm{H}}   {\bf{J}}^{-1}  {\bf{x}}}  \right )$ is jointly convex in $\left\{ {\bf{x}},{\bf{J}}\right\} $, we will prove the inequality in (\ref{kolsdd}) by obtaining a lower bound through Taylor expansion near a feasible point $\left\{ {\widetilde{\bf{x}}},{\widetilde{\bf{J}}}\right\} $. For a convex function  $g\left ( {\bf{X}} \right )$ of matrix ${\bf{X}}$, we have $g\left ( {\bf{X}} \right ) \ge g\left ( {\widetilde{\bf{X}}} \right ) +{\rm{tr}}\left (   \nabla_{\bf{X}} g\left ( {\widetilde{\bf{X}}}  \right )  \left ( {\bf{X}}- {\widetilde{\bf{X}}}\right )    \right )$. Therefore, the lower bound of ${\rm{tr}}\left ( {{\bf{x}}^{\rm{H}}   {\bf{J}}^{-1}  {\bf{x}}}  \right )$ can be obtained as
\begin{equation}\label{koppsd}
	\begin{split} 
		&\ {\rm{tr}}\left ( {{\bf{x}}^{\rm{H}}   {\bf{J}}^{-1}  {\bf{x}}}  \right )  \\
		\ge &\ 2{\rm{Re}}\left (  {\widetilde{\bf{x}}}^{\rm{H}} {\widetilde{\bf{J}}}^{-1} {\bf{x}}\right ) - 2{\rm{Re}}\left (  {\widetilde{\bf{x}}^{\rm{H}}}  {\widetilde{\bf{J}}}^{-1} {\widetilde{\bf{x}}}\right )+{\widetilde{\bf{x}}^{\rm{H}}}  {\widetilde{\bf{J}}}^{-1} {\widetilde{\bf{x}}}\\
		&- {\rm{tr}}\left ({\widetilde{\bf{J}}}^{-1} {\widetilde{\bf{x}}} {\widetilde{\bf{x}}}^{\rm{H}}   {\widetilde{\bf{J}}^{-1}}{\bf{J}} \right ) + {\rm{tr}}\left ({\widetilde{\bf{J}}}^{-1} {\widetilde{\bf{x}}} {\widetilde{\bf{x}}}^{\rm{H}}   {\widetilde{\bf{J}}^{-1}}{\widetilde{\bf{J}}} \right ).   
	\end{split}
\end{equation}
Since ${\bf{J}}$ is a Hermitian matrix in our problem, the constant terms in the right side of (\ref{koppsd}) add up to 0. Furthermore, denoting ${\bf{X}} = \left [{\bf{x}}_1 , {\bf{x}}_2, \cdots,{\bf{x}}_M  \right ] $, we can obtain the following equations with matrix ${\bf{X}}$:
\begin{equation}\label{XJXxjx}
	\begin{split}
	&{\bf{X}}^{\rm{H}}{\bf{J}}^{-1}{\bf{X}}\\
	 =& \begin{bmatrix}
		{\bf{x}}_{1}^{\rm{H}}{\bf{J}}^{-1}{\bf{x}}_{1}    & {\bf{x}}_{1}^{\rm{H}}{\bf{J}}^{-1}{\bf{x}}_{2} & \cdots & {\bf{x}}_{1}^{\rm{H}}{\bf{J}}^{-1}{\bf{x}}_{M}\\
		{\bf{x}}_{2}^{\rm{H}}{\bf{J}}^{-1}{\bf{x}}_{1} & {\bf{x}}_{2}^{\rm{H}}{\bf{J}}^{-1}{\bf{x}}_{2} & \cdots & {\bf{x}}_{2}^{\rm{H}}{\bf{J}}^{-1}{\bf{x}}_{M}\\
		\cdots & \cdots & \cdots & \cdots\\
		{\bf{x}}_{M}^{\rm{H}}{\bf{J}}^{-1}{\bf{x}}_{1}& {\bf{x}}_{M}^{\rm{H}}{\bf{J}}^{-1}{\bf{x}}_{2} & \cdots &{\bf{x}}_{M}^{\rm{H}}{\bf{J}}^{-1}{\bf{x}}_{M}
	\end{bmatrix},
	\end{split}
\end{equation}
\begin{equation}\label{lsjsmd}
	{\rm{tr}}\left ( {\bf{X}}^{\rm{H}}{\bf{J}}^{-1}{\bf{X}} \right ) = \sum_{i=1}^{M} {\bf{x}}_{i}^{\rm{H}}{\bf{J}}^{-1}{\bf{x}}_{i}.
\end{equation}

As we apply the similar approach above, inequality (\ref{koppsd}) can be extended to the form of matrix, namely
\begin{equation}
	\begin{split}
	&{\rm{tr}}\left ( {{\bf{X}}^{\rm{H}}   {\bf{J}}^{-1}  {\bf{X}}}  \right )  \\
	\ge & 2{\rm{Re}}\left ({\rm{tr}}\left(   {\widetilde{\bf{X}}}^{\rm{H}} {\widetilde{\bf{J}}}^{-1} {\bf{X}}\right )\right)  - {\rm{tr}}\left (  {\widetilde{\bf{X}}^{\rm{H}}}  {\widetilde{\bf{J}}}^{-1} {\widetilde{\bf{X}}}\right ).
	\end{split}
\end{equation}

\section{Method of obtaining ${\bf{M}}_{1,1}$  through ${\bf{Q}}_{1}$ }
We have defined ${\bf{G}} = \gamma {\bf{a}}\left ( \theta  \right ) {\bf{a}}^{\rm{H}}\left ( \theta  \right ) $, which is obviously a rank-1 matrix and can be further denoted as ${\bf{G}} =\gamma \left [\lambda_{1}{\bf{a}}\left ( \theta  \right ),\lambda_{2}{\bf{a}}\left ( \theta  \right ),\cdots,\lambda_{M}{\bf{a}}\left ( \theta  \right )  \right ]$, where $\lambda_{i}$ denotes the $i$-th element of ${\bf{a}}^{\rm{H}}\left ( \theta  \right )$ and $\left | \lambda_{i} \right | ^{2} = 1$. Therefore, let ${\bf{F}} = {\bf{H}}_{\rm{BR}}{\bf{E}} {\bf{H}}_{\rm{BR}}^{\rm{H}} \succeq {\bf{0}} $, ${\bf{G}}_{i} = {\rm{diag}}\left(\gamma\lambda_{i}{\bf{a}}\left ( \theta  \right )\right)$, and we have
\begin{equation}\label{maujkl}
	  {\rm{diag}} \left ( {\rm{vec}} \left ( {\bf{G}} \right ) \right )=\begin{bmatrix}
			{\bf{G}}_{1}  & {\bf{0}} & \cdots & {\bf{0}}\\
			{\bf{0}}  &  {\bf{G}}_{2} & \cdots & {\bf{0}}\\
			\cdots  & \cdots & \cdots &\cdots \\
			{\bf{0}}  & {\bf{0}} & \cdots & {\bf{G}}_{N}
		\end{bmatrix}.
\end{equation}

In addition, ${\bf{Q}}_1$ can be expressed as
\begin{equation}\label{matrik}
\begin{split}
	{\bf{Q}}_{1} =&  {\sigma_{v_1}^{2}}{\rm{diag}} \left ( {\rm{vec}} \left ( {\bf{G}} \right ) \right )^{\rm{H}}   \left ( {\bf{I}}_{N} \otimes {\bf{F}}  \right ){\rm{diag}} \left ( {\rm{vec}} \left ( {\bf{G}} \right ) \right )\\
	=&{\sigma_{v_1}^{2}}\begin{bmatrix}
		{\bf{G}}_{1}^{\rm{H}}{\bf{F}}{\bf{G}}_{1}  & {\bf{0}} & \cdots & {\bf{0}}\\
		{\bf{0}}  &  {\bf{G}}_{2}^{\rm{H}}{\bf{F}}{\bf{G}}_{2} & \cdots & {\bf{0}}\\
		\cdots  & \cdots & \cdots &\cdots \\
		{\bf{0}}  & {\bf{0}} & \cdots & {\bf{G}}_{N}^{\rm{H}}{\bf{F}}{\bf{G}}_{N}
	\end{bmatrix}\\
	=&{\sigma_{v_1}^{2}}\begin{bmatrix}
		{\bf{G}}_{1}^{\rm{H}}{\bf{F}}{\bf{G}}_{1}  & {\bf{0}} & \cdots & {\bf{0}}\\
		{\bf{0}}  &  {\bf{G}}_{1}^{\rm{H}}{\bf{F}}{\bf{G}}_{1} & \cdots & {\bf{0}}\\
		\cdots  & \cdots & \cdots &\cdots \\
		{\bf{0}}  & {\bf{0}} & \cdots & {\bf{G}}_{1}^{\rm{H}}{\bf{F}}{\bf{G}}_{1}
	\end{bmatrix}.
\end{split}
\end{equation}
Note that ${\bf{G}}_{1}^{\rm{H}}{\bf{F}}{\bf{G}}_{1} \succeq {\bf{0}}$, it can be easily derived that ${\bf{Q}}_{1}$ is also positive semi-definite. To tackle the problem of obtaining ${\bf{M}}_{1,1}$ through ${\bf{Q}}_{1}$, equation ${\rm{tr}}\left ( {\bf{ABCD}} \right )  = \left (  {\rm{vec}}\left ( {\bf{D}}^{\rm{T}} \right ) \right ) ^{\rm{T}}  \left ({\bf{C}}^{\rm{T}} \otimes {\bf{A}} \right )  {\rm{vec}}\left ( {\bf{B}} \right )$ is applied again. Let ${\bf{A}} = {\bf{G}}_{1}^{\rm{H}}{\bf{F}}{\bf{G}}_{1}$, ${\bf{B}} = {\bf{D}}={\bf{V}}$ and ${\bf{C}} ={\bf{I}}_{N}$, then we have 
\begin{equation}\label{accerf}
	\begin{split}
		&{\hat{\bf{v}}}^{\rm{H}}{\bf{Q}}_{1}{\hat{\bf{v}}} \\
		=&\left ( {\rm{vec\left ( {\bf{V}} \right ) }} \right ) ^{\rm{H}}\left ({\bf{I}}_{N} \otimes\left( {\bf{G}}_{1}^{\rm{H}}{\bf{F}}{\bf{G}}_{1}\right) \right ) {\rm{vec\left ( {\bf{V}} \right ) }}\\
		=&\left ( {\rm{vec}}\left ( {\bf{V}}^{\rm{T}} \right ) \right )  ^{\rm{T}}\left ({\bf{I}}_{N} \otimes\left( {\bf{G}}_{1}^{\rm{H}}{\bf{F}}{\bf{G}}_{1}\right) \right ) {\rm{vec\left ( {\bf{V}} \right ) }}\\
		=&{\rm{tr}}\left ( {\bf{G}}_{1}^{\rm{H}}{\bf{F}}{\bf{G}}_{1}{\bf{V}}^{\rm{H}}{\bf{V}} \right ) \\
		=&{\rm{tr}}\left ( {\bf{V}}{\bf{M}}_{1,1}{\bf{V}}^{\rm{H}} \right ),
	\end{split}
\end{equation}
where matrix ${\bf{M}}_{1,1}$ is actually ${\bf{G}}_{1}^{\rm{H}}{\bf{F}}{\bf{G}}_{1}$, and can be obtained by equation (\ref{aserjk}) through matrix ${\bf{Q}}_{1}$ and have been proved to be positive semi-definite. Since ${\bf{G}} = \gamma {\bf{a}}\left ( \theta  \right ) {\bf{a}}^{\rm{H}}\left ( \theta  \right ) $, it is easy to derive that ${\rm{rank}}\left ( {\bf{G}} \right ) = 1$, and the ranks of ${\bf{A}} = {\bf{H}}^{\rm{H}}_{\rm{BR}}{\bf\Phi}^{\rm{H}} {\bf{G}}{\bf\Phi}{\bf{H}}_{\rm{BR}}$, ${\bf{E}} ={{\bf{J}}}^{-1}_{(l)}{ {\bf{A}}} _{(l)} {\bf{R}} {{\bf{A}}}^{\rm{H}}_{(l)} {{\bf{J}}}^{-1}_{(l)}$ , ${\bf{F}} = {\bf{H}}_{\rm{BR}}{\bf{E}} {\bf{H}}_{\rm{BR}}^{\rm{H}} $ and ${\bf{M}}_{1,1} = {\bf{G}}_{1}^{\rm{H}}{\bf{F}}{\bf{G}}_{1}$ all turn to be 1. The positive semi-definiteness of ${\bf{Q}}_{2}, {\bf{Q}}_{3}$ and the method of obtaining ${\bf{M}}_{1,2}, {\bf{M}}_{2}$ can be derived likewise.
\end{appendices}

\bibliographystyle{IEEEtran}
\bibliography{20231201}

\end{document}